\documentclass[showpacs,preprintnumbers,twocolumn,aps,prl,reprint,superscriptaddress,nobibnotes,nofootinbib]{revtex4-1}

\usepackage{amsmath,amssymb,hyperref,times,graphicx,bm,bbold}
\usepackage{bibunits}
\usepackage{subfig}
\usepackage[export]{adjustbox}
 \usepackage{enumitem}

\usepackage{physics}

\newcommand{\be}{\begin{equation}}
\newcommand{\ee}{\end{equation}}
\newcommand{\bea}{\begin{eqnarray}}
\newcommand{\eea}{\end{eqnarray}}

\def\up{\uparrow}
\def\dn{\downarrow}
\def\<{\langle}
\def\>{\rangle}

\providecommand{\up}{\uparrow}
\providecommand{\dn}{\downarrow}

\newcommand{\para}[1]{\left(#1\right)}

\newcommand{\COMMENT}[1]{}

\begin{document}

\defaultbibliographystyle{apsrev4-1_custom}
\defaultbibliography{francesco,extra}

\title{Entanglement Hamiltonian of Interacting Systems: Local Temperature Approximation and Beyond}

\author{\firstname{Mahdieh} \surname{Pourjafarabadi}}
\email{The two first authors have contributed equally to this work.}
\affiliation{\mbox{Department of Physics, Sharif University of Technology, Tehran 14588-89694, Iran}}
\author{\firstname{Hanieh} \surname{Najafzadeh}}
\email{The two first authors have contributed equally to this work.}
\affiliation{\mbox{Department of Physics, Sharif University of Technology, Tehran 14588-89694, Iran}}
\author{\firstname{Mohammad-Sadegh} \surname{Vaezi}}
\affiliation{\mbox{Pasargad Institute for Advanced Innovative Solutions (PIAIS) , Tehran 19916-33361, Iran}}
\author{\firstname{Abolhassan} \surname{Vaezi}}
\email{Corresponding author. Email address: vaezi@sharif.edu}
\affiliation{\mbox{Department of Physics, Sharif University of Technology, Tehran 14588-89694, Iran}}

\begin{abstract}
We investigate the second quantization form of the entanglement Hamiltonian (EH) of various subregions for the ground-state of several interacting lattice fermions and spin models. The relation between the EH and the model Hamiltonian itself is an unsolved problem for the ground-state of generic local Hamiltonians. In this article, we demonstrate that the EH is practically local and its dominant components are related to the terms present in the model Hamiltonian up to a smooth spatially varying temperature even for (a) discrete lattice systems, (b) systems with no emergent conformal or Lorentz symmetry, and (c) for subsystems with non-flat boundaries, up to relatively strong interactions. We show that the mentioned local temperature at a given point decays inversely proportional to its distance from the boundary between the subsystem and the environment. We find the subdominant terms in the EH as well and show that they are severely suppressed away from the boundaries of subsystem and are relatively small near them.
\end{abstract}

\maketitle
\begin{bibunit}
{\it Introduction.}---
Entanglement is a unique feature of quantum mechanics and serves as an essential tool in quantum information, quantum gravity, identification of topological order, quantum phase transition, etc \cite{susskind2005introduction,nielsen2002quantum,susskind2005introduction,ryu2006holographic,kitaev2006topological,levin2006detecting,li2008entanglement,sterdyniak2012real,PhysRevLett.104.180502,lauchli2010disentangling,liu2015non,cian2020engineering,vaezi2017numerical,zaletel2013topological,qi2012general,pollmann2010entanglement,yarloo2018anyonic,calabrese2009entanglement,pichler2016measurement}. The entanglement Hamiltonian (EH) associated with a subregion $A$ embedded in a manifold $M=A \cup B$ is defined as $\rho_A= e^{-K_A}$. Here $\rho_A = \mathrm{Tr}_{B} \rho_{M}$ denotes the reduced density matrix (RDM) of $A$, where $\rho_M$ represents the total density matrix. 
One important question that arises from this definition is the relation between $K_A$ and $H_A$, the Hamiltonian terms with support only in region $A$. In fact, this problem dates back to the 19th century. A cornerstone of the classical statistical mechanics is that a subsystem $A$ at thermal equilibrium with its environment ($B$) is described by a thermal ensemble with $K_A = {H_A}/{T_0}$ ($k_B=\hbar=c=1$) where $T_0$ is a uniform and position independent temperature. Furthermore, the eigenstate thermalization hypothesis conjectures that the RDM of highly excited quantum states will look thermal, again with $K_A={H_A}/{T_0}$, where the uniform temperature $T_0$ in this case is dictated by the energy density \cite{d2016quantum}. In this article, we revisit this fundamental problem and using the density-matrix-renormalization-group (DMRG) approach we obtain the second quantization form of $K_A$ for a number of interacting model Hamiltonians and for a variety of boundary shapes and conditions. Comparing the components of $K_A$ and $H_A$, we demonstrate that the above mentioned statements are not quite accurate and for the ground-state of local Hamiltonians, $K_A$ is indeed well-approximated by a {\em local} and {\em non-uniform temperature} rather than a uniform one. 

The theoretical form of $K_A$ is known only for a limited class of {\em continuum} models with {\it conformal} symmetry (or {\it Lorentz} symmetry at zero temperature) and only for certain geometries of $A$ (e.g., {\it half-space} or {\it ball} geometry). It is known that under these conditions: (i) $K_A$ is  local, (ii) the EH density is related to the Hamiltonian density via a smooth local temperature, namely $K_A = \int_{x\in A} d^d x~ \mathcal{K}_A(x)  = \int_{x\in A} d^d x \frac{\mathcal{H}\para{x}}{T\para{x}}$, and (iii) $T(x)$ approaches $T_0$, the equilibrium temperature of the entire system, far away from $\partial A$ (the boundary of A) and grows as $\frac{v}{2\pi r(x)}$ at distance $r$ near $\partial A$. Here, $v$ is the group velocity of low energy excitations \cite{bisognano1975duality,bisognano1976duality,susskind2005introduction,casini2011towards,cardy2016entanglement,arias2017local,bousso2015entropy,vaezi2018locala,vaezi2018localb,turkeshi2019entanglement}. We refer to these findings as the local temperature approximation (LTA)~\cite{vaezi2018locala,vaezi2018localb}. 

The LTA can be justified using the following intuitive argument. In thermal systems, the entropy density is proportional to their temperatures. On the other hand, for ground-states, instead of the thermal entropy, we deal with the entanglement entropy which is not precisely an extensive property. Nevertheless, we can still consider and gauge the contribution of individual degrees of freedom residing inside $A$ to the overall entanglement entropy between $A$ and $B$, $S_A$. Indeed, quantum mutual information can be one candidate to quantify such local contributions. Due to the decay of quantum mutual information with distance for the ground-state of local Hamiltonians, the degrees of freedom that live near $\partial A$, are more entangled with those residing at $B$ than more distant ones. Accordingly, we can assign an {\em effective quantum local temperature} to different subregions of $A$ proportional to their contributions to $S_A$, which as just discussed must diminish away from $\partial A$.

\begin{figure}
\setlength{\belowcaptionskip}{-5pt}
\includegraphics[width=7.0cm]{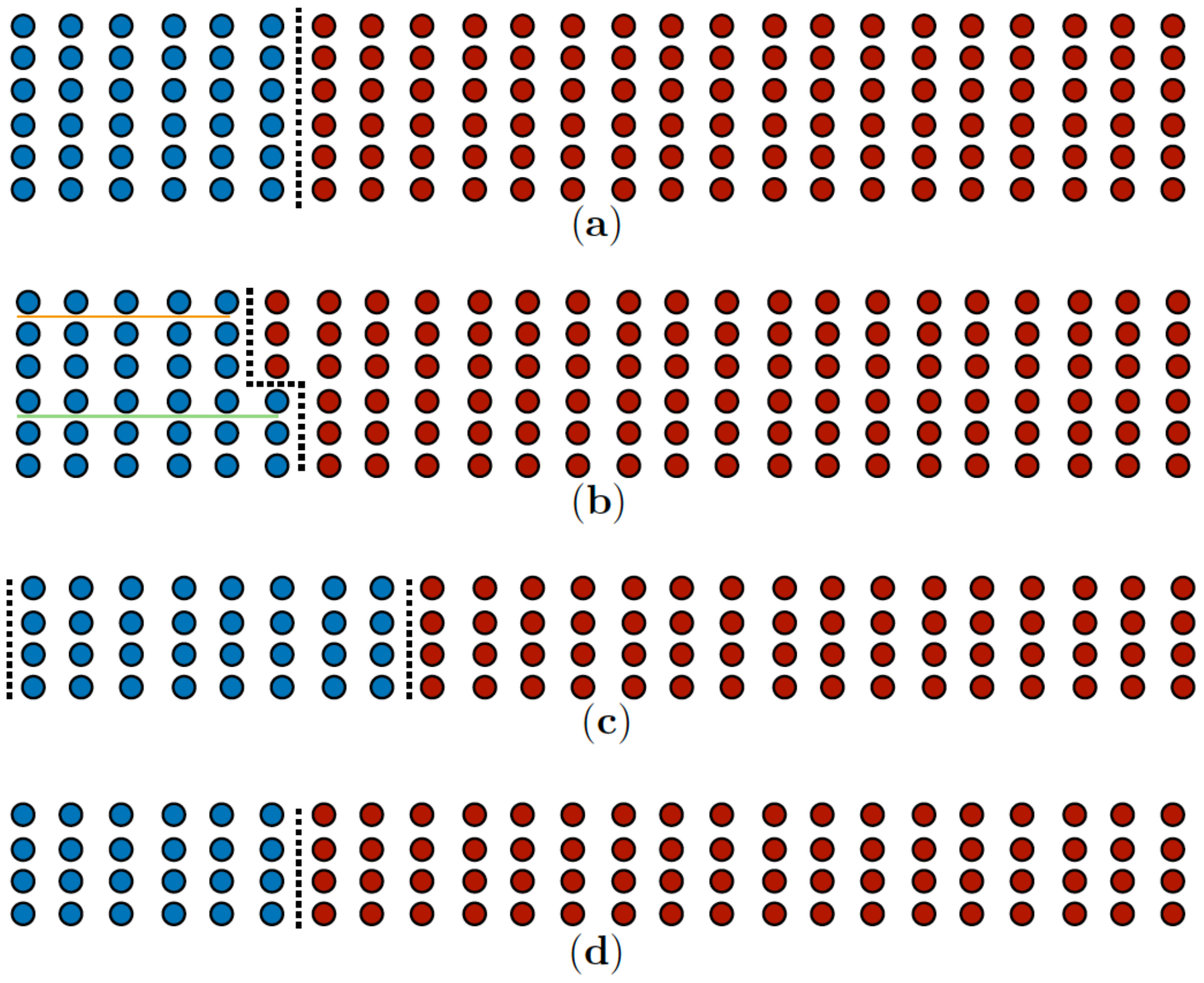}
\caption{\raggedright Various geometries of system and subsystem that we consider in this article for computing the EH. The blue (red) sites define the subsystem $A$ ($B$). The dotted lines indicate the boundary between $A$ and $B$, $\partial A$. In (c), $\partial A$ comprises two disjoint surfaces since we have considered torus geometry. The green and orange lines in (b) illustrate the rows at which the couplings in Fig.~\ref{fig2} are plotted respectively.}
\label{fig0}
\end{figure}

Numerically, the EH of free fermions and free bosons can be evaluated easily \cite{chung2001density,cheong2004many,peschel2009reduced}. However, for interacting models, it becomes highly nontrivial and challenging. Recently, several studies have analyzed the EH of quasi-one-dimensional conformal invariant or integrable models \cite{toldin2018entanglement,dalmonte2018quantum,giudici2018entanglement,eisler2019continuum,mendes2019entanglement,wong2013entanglement,zhu2019reconstructing,furukawa2011entanglement,lauchli2012entanglement,chen2013quantum,wen2018entanglement,koeller2018local,lashkari2016modular,di2020entanglement,zhang2020lattice,zhu2020entanglement,kokail2020entanglement}. Nonetheless, we are still lacking a systematic derivation of the EH for larger and generic interacting systems and for various boundary geometries. In this article, we address this problem and introduce a DMRG-based algorithm that enables us to extract the EH for a broader spectrum of problems.

\noindent{\bf ${\bf J_1-J_2}$ Heisenberg model.--} Let us first discuss the form of $K_A$ for the $J_1-J_2$ Heisenberg model ($J_1 = 1$) with the following Hamiltonian on the square lattice:

\bea
H = J_1 \sum_{\left<{\bf ij}\right> \in M} {\bf S}_{\bf i}.{\bf S}_{\bf j} +J_2 \sum_{\left<\left<{\bf ij}\right>\right> \in M} {\bf S}_{\bf i}.{\bf S}_{\bf j}.   \label{H1}
\eea
The above Hamiltonian respects a SU(2) symmetry. Hence, $K_A$ must respect SU(2) symmetry as well and thus expanded as follows:
\bea
K_A = \sum_{\bf i,j\in A} g_{J,\bf ij} \, {\bf S}_{\bf i}.{\bf S}_{\bf j} + \cdots~. \label{EH1}
\eea
Note that there is no restriction on ${\bf i} =\para{i_x,i_y}$ and ${\bf j} =\para{j_x,j_y}$ except that both must belong to $A$. In our study, we have dropped higher order terms since the retained terms already yield satisfactory results. 

The DMRG technique is based on identifying the most relevant basis states of the Hilbert space \cite{white1992density}. Then we truncate the Hilbert space and discard the less relevant states. The number of kept states which controls the accuracy of DMRG is called the bond dimension, $\chi$, and its default value equals $2^{10}$ throughout this article. The procedure of finding the truncation operators consecutively involves the computation and diagonalization of the RDM at every step of DMRG and for different subsystem sizes. Hence, $\rho_A$ is a natural byproduct of DMRG method and is available at every step. Moreover, every operator component of $K_A$ (e.g., ${\bf S_i. S_j}$) has a matrix representation in DMRG, albeit in the truncated subspace. The remaining task is to adjust the EH's couplings, $g_{J,\bf ij}$, to bring our (simplified) guess for $K_{A}$, which we will denote as $\widetilde{K_A}$, close enough to the matrix representation of the RDM in the truncated Hilbert space achieved via DMRG: $K_A = -\log \rho_A$. To this end, we need to define an appropriate cost function as a measure of the distance between $\widetilde{K_A}$ and $K_A$. In our investigations, we mainly utilized the Hilbert-Schmidt distance between the Green's functions, namely $\Delta_1 := {\rm Tr}_A\para{G_A-\widetilde{G}_A}^2$, where $G_{A,\bf ij}={\rm Tr}_A\para{{\bf S_i. S_j} {\rho}_A}$ is the Green's function matrix achieved by DMRG for $A$  and $\widetilde{G_A}$ denotes its counterpart evaluated using the trial RDM, $\widetilde{\rho_A}=\exp(-\widetilde{K_A})$
~\cite{swingle2014reconstructing}. This cost function yields more reliable and reasonably robust results (against changing $\chi$) in the truncated Hilbert space than other candidates, e.g., the quantum relative entropy between $\rho_A$ and $\widetilde{\rho_A}$. We find the latter to overfit to numerical noises, e.g., the truncation and computer's roundoff errors (see Appendix D for more details). In the optimization procedure, we initialized $g_{J,\bf ij}$ based on general expectations from LTA, e.g., the locality of $g_{J,\bf ij}$ and its linear dependence on $x_{\bf ij} $ (the minimum distance between $\overline{\bf ij} = \frac{\bf i+j}{2}$ and $\partial A$). Then, we employed the gradient descent algorithm and let the cost function to decide the optimum choice for $g_{J,\bf ij}$ (see the Appendix for more details). 

We first focus on $J_2 = 0$ Heisenberg model which is unfrustrated and is known to host a Néel order on the square lattice \cite{sandvik1997finite,stoudenmire2012studying}. Thus, its ground-state is a symmetry broken phase with gapless Goldstone modes and does not respect the full conformal symmetry (e.g., the translational and (around the center of plaquettes) rotational symmetries are broken). For this model, the system is always subject to the PBC along $y$ axis.

\begin{itemize}[wide]
\item As the first example, we study the manifold and subsystem $A$ depicted in Fig.~\ref{fig0}-a, where an OBC is imposed along $x$. 
In this case, $\partial A$ is flat and its locus is given by $x_b = 6 + 1/2$, the line which splits columns 6 and 7. The optimum couplings, $g_{J,\bf ij}$, which reproduce the DMRG's Green's functions (with less than 0.1 $\%$ error), are plotted in Fig.~\ref{fig1}.

Fig.~\ref{fig1a} shows the nearest neighbor (NN) couplings along $x$ and $y$ (more precisely, $\beta_{J,x}\para{i_x+1/2} := g_{J,\bf i,i+\hat{x}}$ and $\beta_{J,y}\para{i_x} := g_{J,\bf i,i+\hat{y}}$) which are independent of $i_y$ due to the $y$-axis translation preserving shape of $A$. Indeed, $\beta_J$'s are the inverse local temperature profiles. As Fig.~\ref{fig1a} suggests, $\beta_{J,x}$ and $\beta_{J,y}$ follow the same profile, albeit if we shift the argument of $\beta_x$ by half of the lattice spacing. This shift is due to the fact that for $g_{J,x}$, the start and end points are located at different positions along $x$, while for $g_{J,y}$, the two points have identical $x$ values. In Appendix D, we demonstrate the robustness of $\beta_{J,x}$ and $\beta_{J,y}$ versus $\chi$.
In Fig.~\ref{fig1b}, we have plotted $g_{J,xy}\para{i_x+1/2} :=  g_{J,\bf i,i+\hat{x}+\hat{y}}$, $g_{J,yy}\para{i_x} :=  g_{J,\bf i,i+2\hat{y}}$, and $g_{J,xx}\para{i_x+1} :=  g_{J,\bf i,i+2\hat{x}}$. Their values are negligible everywhere and they all die off quickly away from $\partial A$.  These imply the locality of $K_A$ for the Heisenberg model when $\partial A$ is flat.

\begin{figure}[t]
\setlength{\belowcaptionskip}{-5pt}
\subfloat[]{%
  \includegraphics[width=4.54cm]{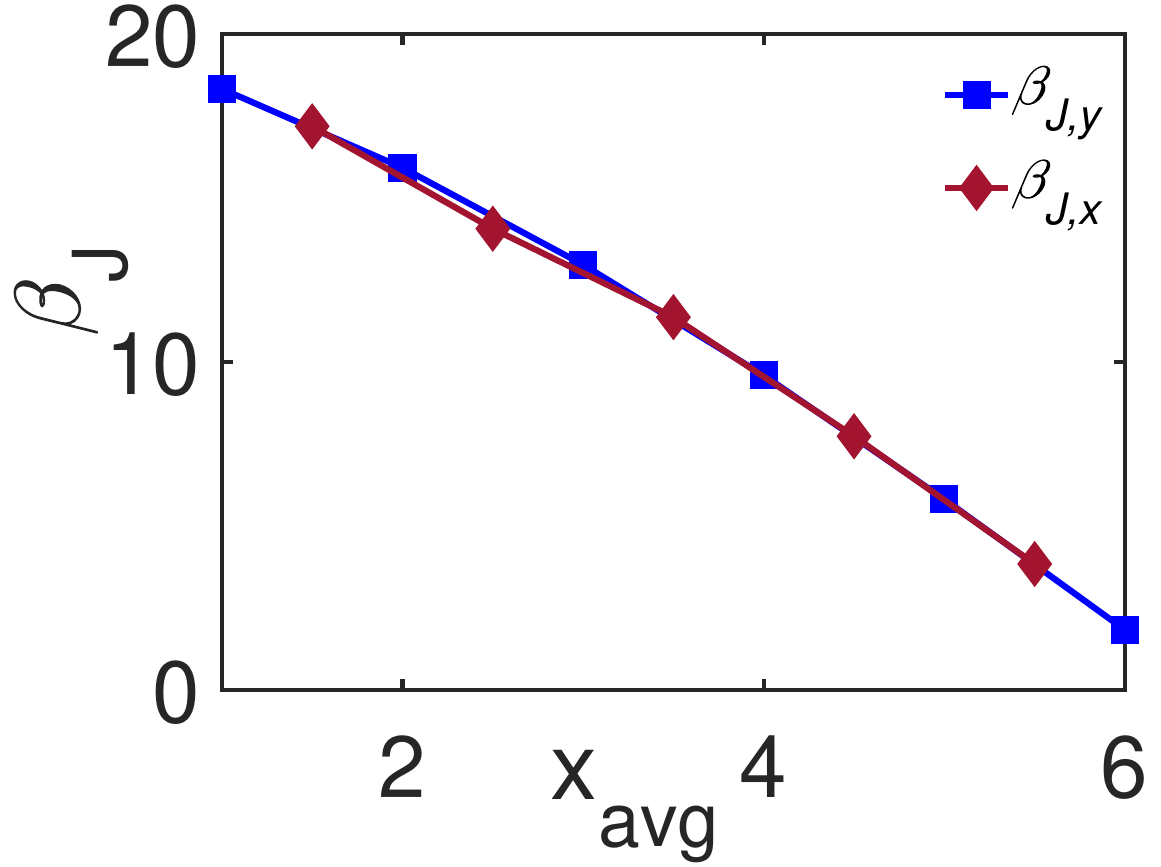}%
  \label{fig1a}%
}
\subfloat[]{%
  \includegraphics[width=4.54cm]{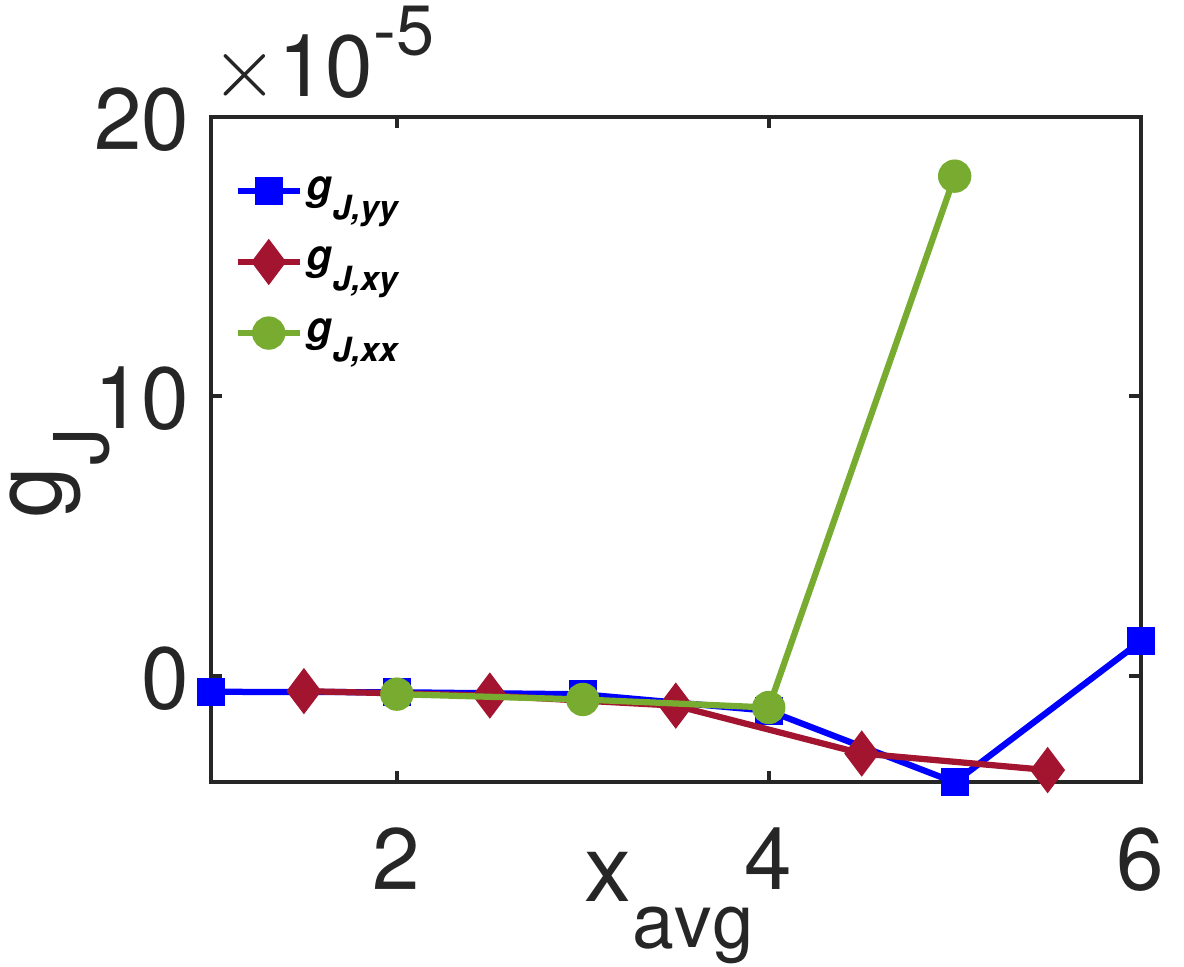}%
  \label{fig1b}%
}
\caption{\raggedright EH's couplings for the Heisenberg model ($J_2=0$) for the subsystem geometry shown in Fig~\ref{fig0}-a. (a) The inverse temperature profiles $\beta_{J,a}(x_{\rm avg})=J_1^{-1}g_{J,a}\para{\bf i,j}$ ($a=x,y$) for the NN couplings versus the midpoint argument $x_{\rm avg}:=\frac{1}{2}\para{i_x+j_x}$. (b) Second and third neighbor couplings versus $x_{\rm avg}$.}
\label{fig1}
\end{figure}

\item We now consider the same conditions as above, but this time with a curved $\partial A$ as shown in Fig.~\ref{fig0}-b. In this case, $\beta_{J,x}$ and $\beta_{J,y}$ will depend on both $i_x$ and $i_y$. In Fig.~\ref{fig2}, we have plotted $\beta_{J,x}$ and $\beta_{J,y}$ for two different rows marked by orange and green lines. Interestingly, $\beta_{J,x}$ and $\beta_{J,y}$ profiles display a somewhat smooth curve satisfying our expectations from LTA. The position dependence of the inverse temperature profile is more complicated in this problem, since the distance between $\bf \overline{ij}=\para{i+j}/2$ and $\partial A$ depends on both its $x$ and $y$ components.  In Figs.~\ref{fig2b} and \ref{fig2d} the second and third neighbor couplings are plotted for the above mentioned rows. We see that for curved boundaries between $A$ and $B$, $K_A$ remains local everywhere, except close to $\partial A$ where we observe additional terms though relatively small and subdominant.

\begin{figure}[t]
\centering
\setlength{\belowcaptionskip}{-5pt}
\subfloat[]{%
  \includegraphics[width=4.54cm]{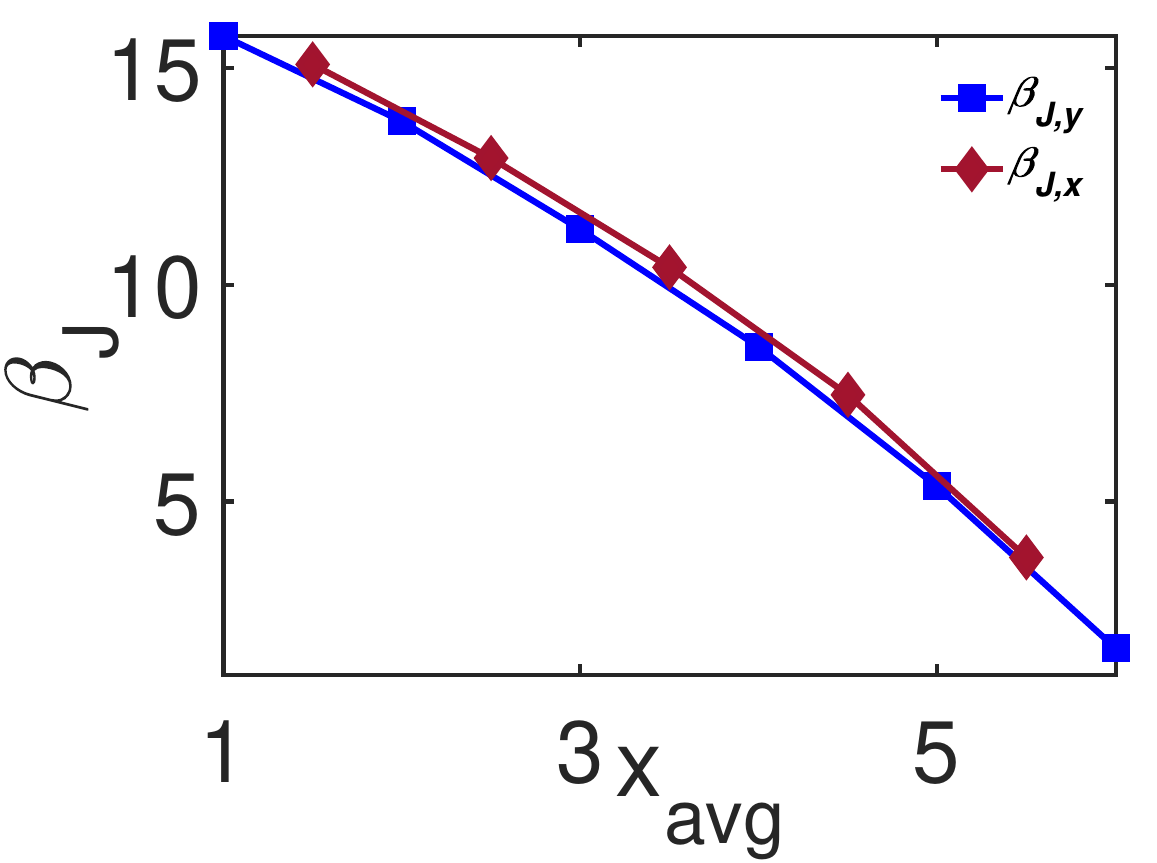}%
  \label{fig2a}%
}
\subfloat[]{%
  \includegraphics[width=4.54cm]{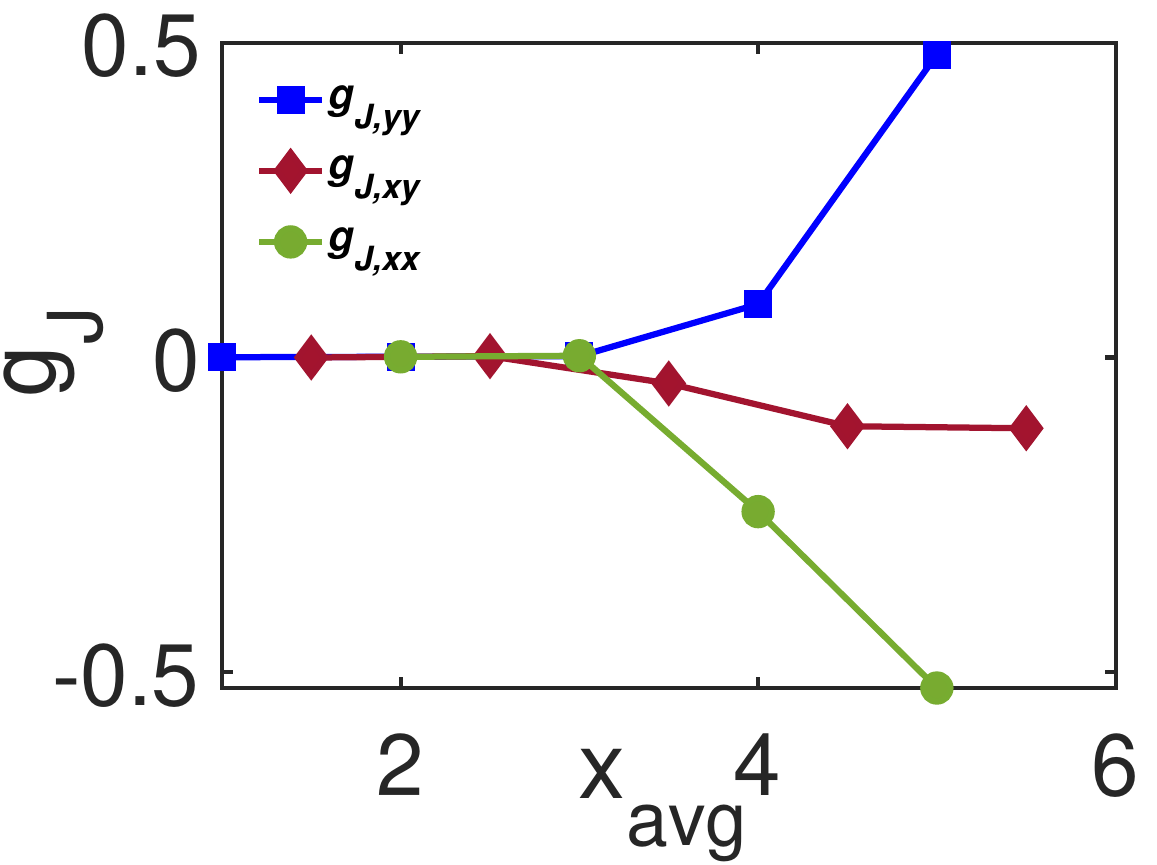}%
  \label{fig2b}%
}\qquad
\subfloat[]{%
  \includegraphics[width=4.54cm]{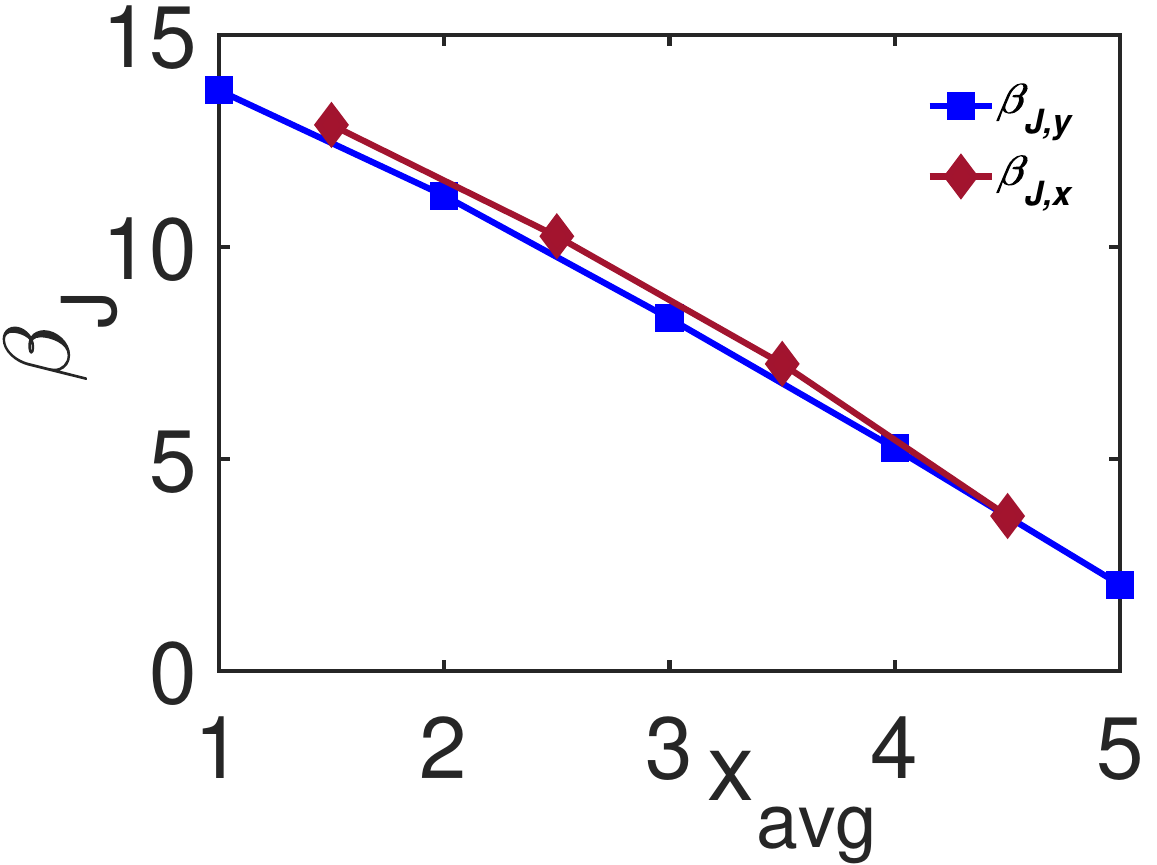}%
  \label{fig2c}%
}
\subfloat[]{%
  \includegraphics[width=4.54cm]{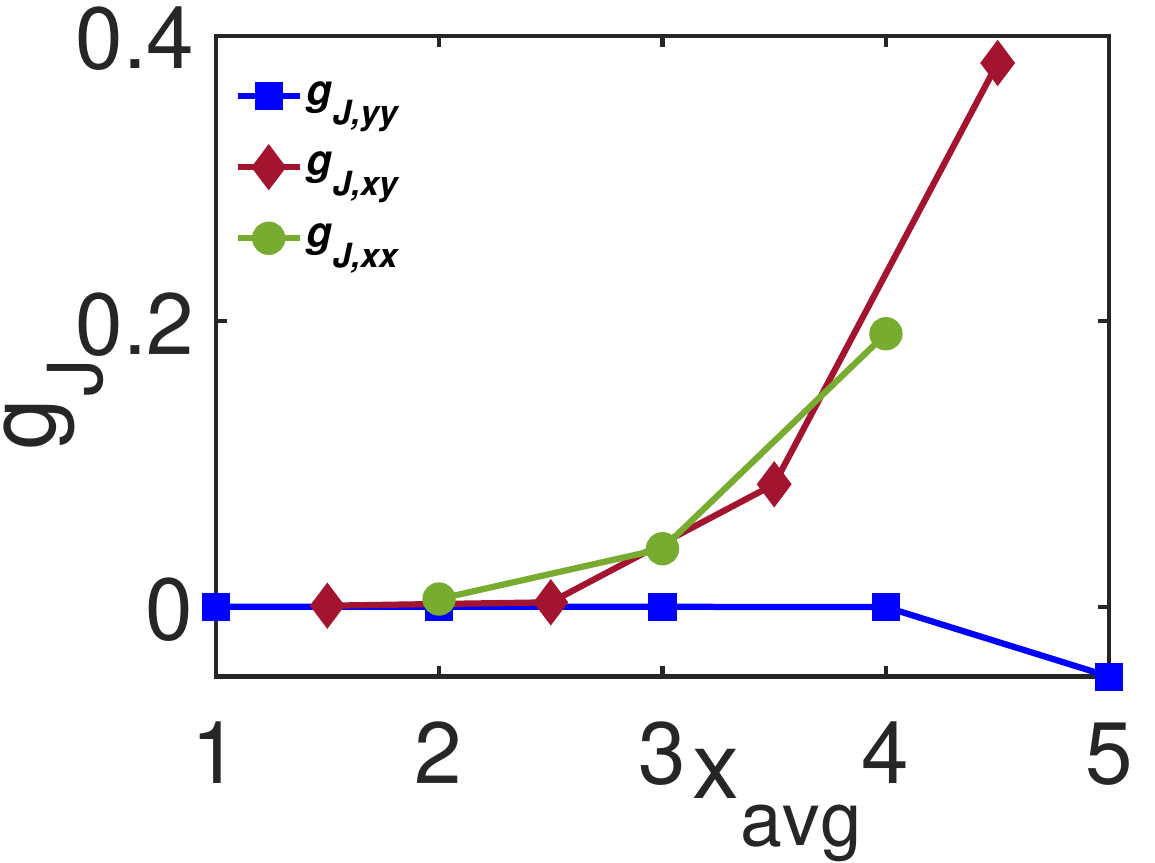}%
  \label{fig2d}%
}
\caption{\raggedright EH's couplings for the Heisenberg model ($J_2=0$) for the subsystem geometry shown in Fig.~\ref{fig0}-b. Since $A$ breaks the translational symmetry along $y$, $g_{J,\para{\bf i,j}}$ depends on both $i_x$ and $j_y$. In (a) and (b), the couplings along the green line in Fig.~\ref{fig0}-b are plotted and in (c) and (d), those corresponding to the orange line. The terms beyond LTA grow substantially near $\partial A$ for non-flat boundary geometries compared to flat boundaries (cf. Fig. \ref{fig1}).}
\label{fig2}
\end{figure}

\item Now, we consider the geometry illustrated in Fig.~\ref{fig0}-c. Since the PBC is imposed on $M$ along both $x$ and $y$ directions, we have chosen $N_y=4$ to ensure $\chi = 2^{10}$ is sufficient for DMRG's convergence. As we see in Fig.~\ref{fig0}-c, $\partial A$ is described by two surfaces, one of them separates columns 6 and 7 and the other one lies between the first and last columns. As a result, LTA predicts that $\beta_{J,x}$ and $\beta_{J,y}$ must follow a parabolic form and vanish near both boundary surfaces. Fig.~\ref{fig3}, shows our numerical results for the nearest as well as further neighbor couplings, both consistent with LTA.

\begin{figure}[t]
\centering
\setlength{\belowcaptionskip}{-5pt}
\subfloat[]{%
  \includegraphics[width=4.54cm]{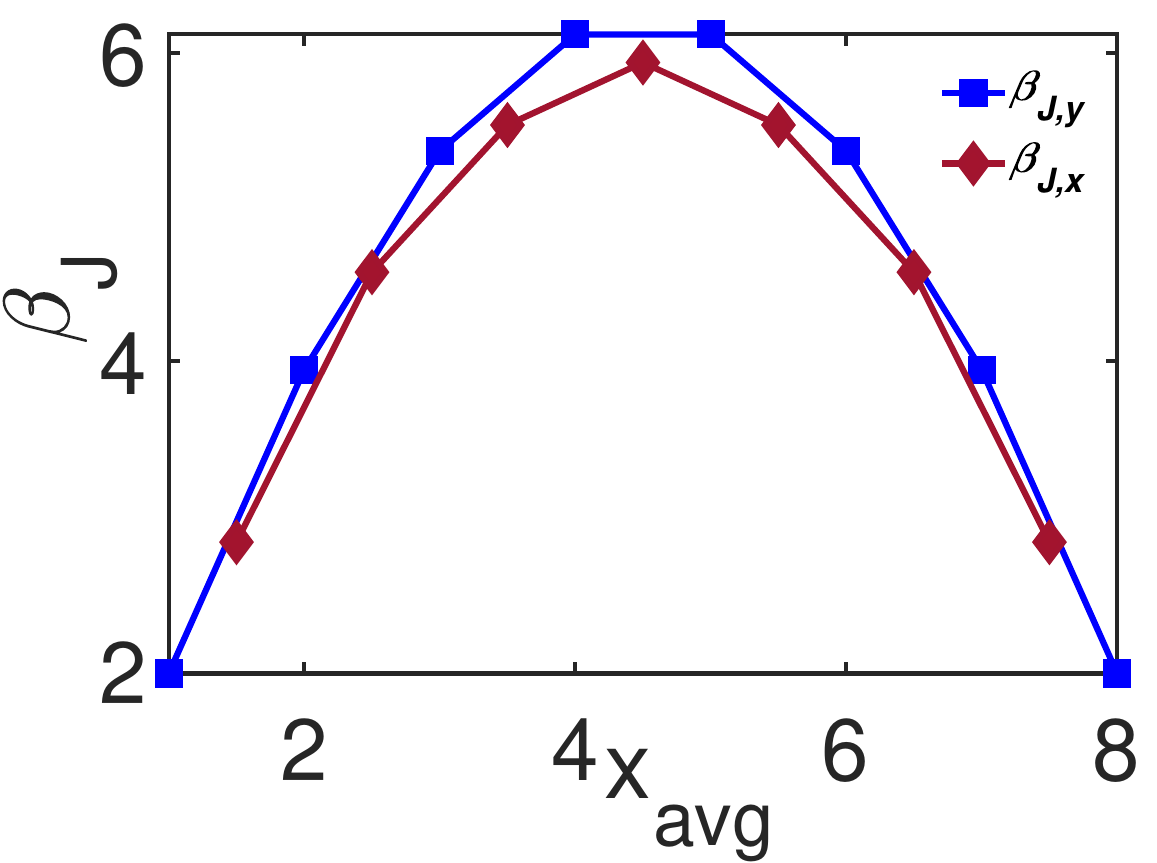}%
  \label{fig:evaluation:revenue}%
}
\subfloat[]{%
  \includegraphics[width=4.54cm]{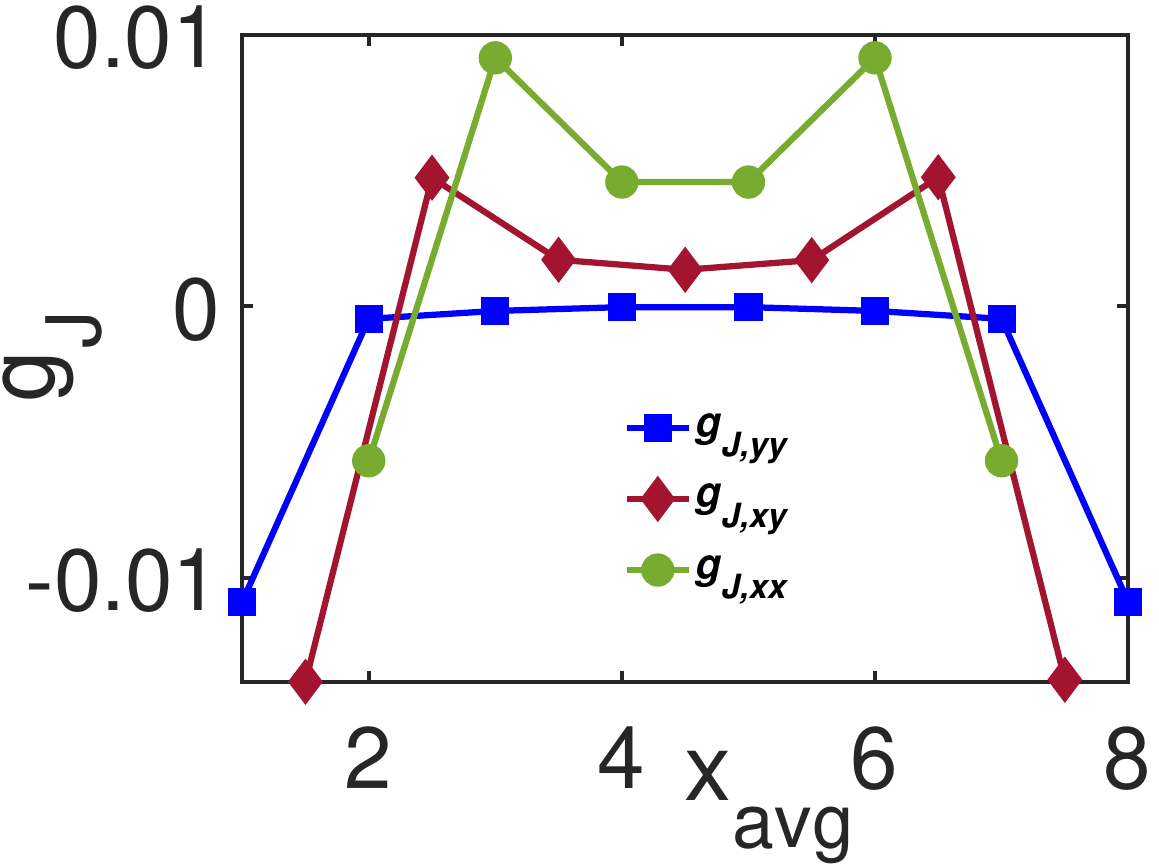}%
  \label{fig3a}%
}
\caption{\raggedright EH's couplings for the Heisenberg model ($J_2=0$) subject to the PBC along $x$, for the subsystem geometry shown in Fig.~\ref{fig0}-c. The NN couplings follow a parabolic curve and die off near both boundaries.}
\label{fig3}
\end{figure}

\item Let us now turn to the frustrated Heisenberg model with $J_2 = 0.6$ whose true ground-state is not well-understood, though it is conjectured to be a spin liquid phase with no classical spin order and algebraically decaying spin-spin correlations \cite{jiang2012spin}. For this model, we consider the geometry depicted in Fig.~\ref{fig0}-d. The ground-state is expected to be more entangled when $J_2/J_1\sim O(1)$. Hence, we consider $N_y=4$ (again $N_x = 24$) to ensure that the ground-state is achieved reliably via $\chi = 2^{10}$ in DMRG. Since the Hamiltonian contains next nearest neighbor (NNN) couplings, we expect significant values for the NNN in $g_{J,\bf ij}$ as well. In Fig.~\ref{fig4a}, $\beta_{J,x}$, $\beta_{J,y}$, and also $\beta_{J,xy}\para{i_x+1/2} := \frac{1}{J_2}g_{J,\bf i,i+\hat{x}+\hat{y}}$ are plotted. Again, we see that all these $\beta_J$'s follow the same curve. Furthermore, Fig.~\ref{fig4b} verifies the locality of $K_A$ everywhere except at  $\partial A$ where $g_{J,yy}$ is about $29\%$ ($28\%$) of $g_{J,y}$ ($g_{J,xy}$) at that location.

\begin{figure}[t]
\centering
\setlength{\belowcaptionskip}{-5pt}
\subfloat[]{%
  \includegraphics[width=4.54cm]{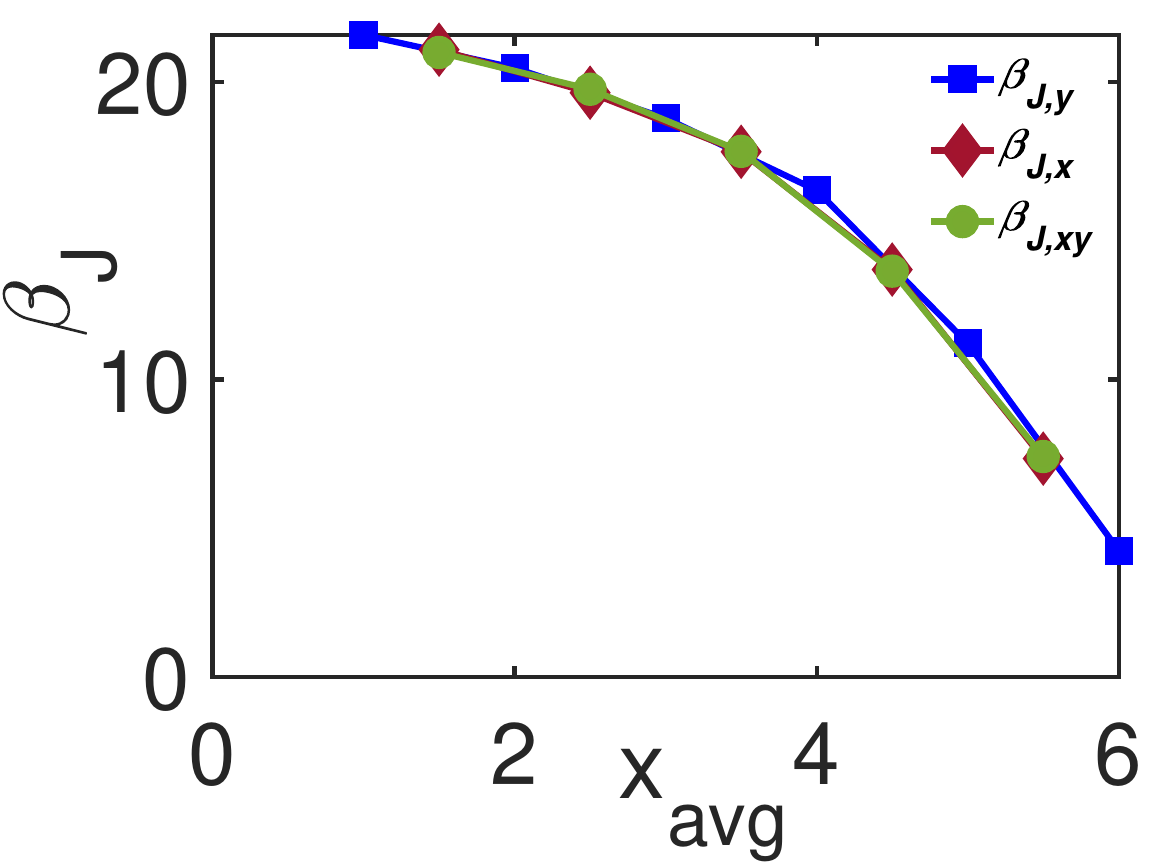}%
  \label{fig4a}%
}
\subfloat[]{%
  \includegraphics[width=4.54cm]{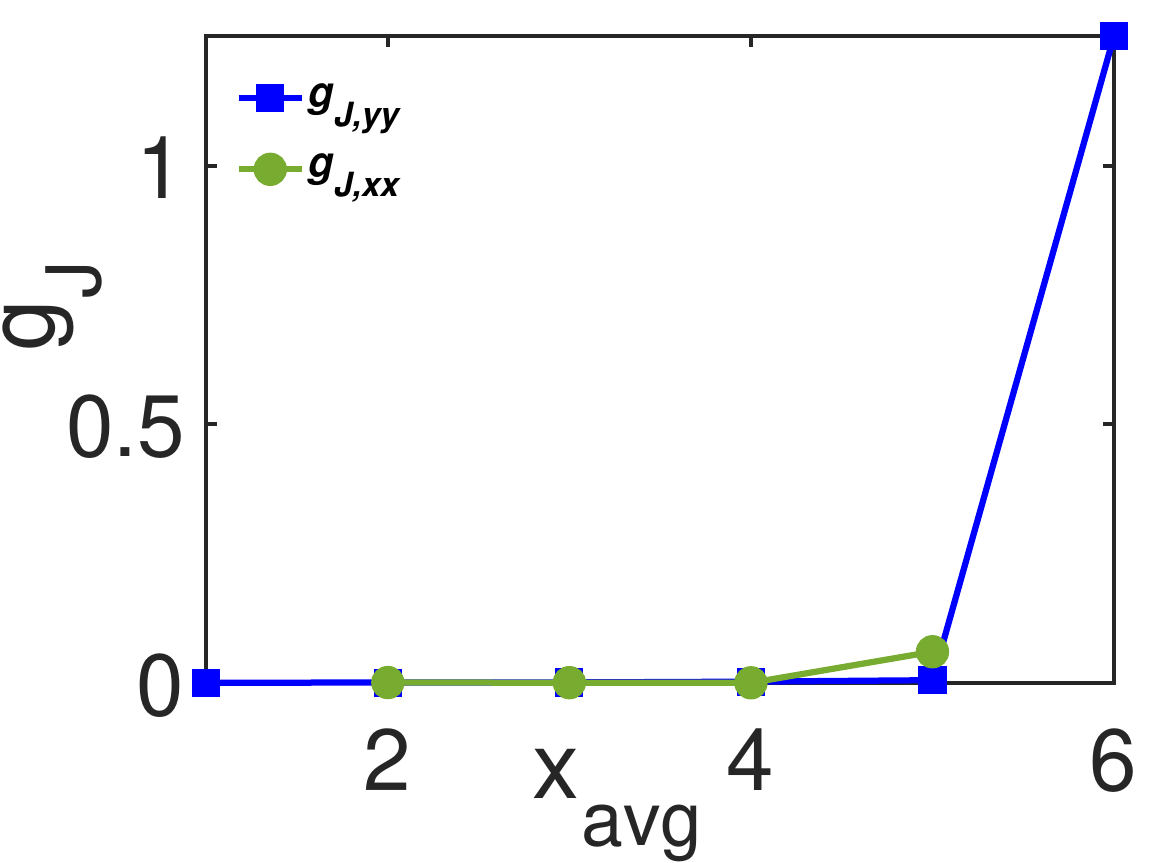}%
  \label{fig4b}%
}
\caption{\raggedright EH's couplings for the $J_1-J_2$ Heisenberg model ($J_2=0.6$) for the subsystem geometry shown in Fig.~\ref{fig0}-d. LTA is valid everywhere, except at the boundary where the second neighbor coupling along $y$ is non-negligible.}
\label{fig4}
\end{figure}

\noindent {\bf Hubbard model.-} Here, we discuss the second quantization form of $K_A$ for the Hubbard model on the square lattice, whose Hamiltonian is:
\bea
H = &&-t_1 \sum_{\left<{\bf ij}\right> \in M,\sigma} c_{\bf i,\sigma}^\dag c_{\bf j,\sigma} -\mu\sum_{{\bf i}\in M,\sigma} n_{\bf i,\sigma}\cr
 &&+U\sum_{{\bf i}\in M}\para{n_{\bf i,\up}-\frac{1}{2}}\para{n_{\bf i,\dn}-\frac{1}{2}},~~~\label{H2}
\eea
where $n_{\bf i,\sigma} = c_{\bf i,\sigma}^\dag c_{\bf i,\sigma}$. In this article, we consider $t_1 = 1$, and $U=4$. The above Hamiltonian enjoys a $U(1) \times SU(2)$ symmetry for generic fillings. Accordingly, $K_A$ must be expanded as follows:
\begin{figure}[t]
\centering
\setlength{\belowcaptionskip}{-5pt}
\subfloat[]{%
  \includegraphics[width=4.54cm]{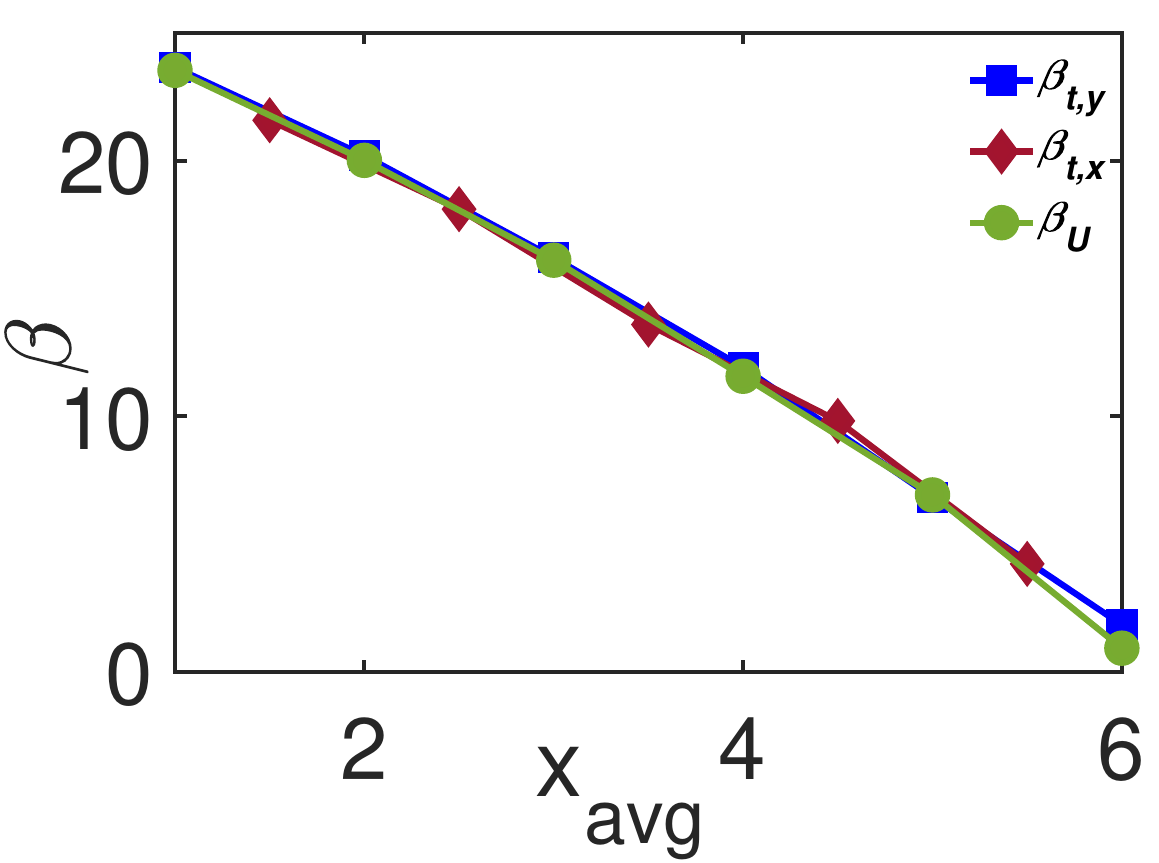}%
  \label{fig5a}%
}
\subfloat[]{%
  \includegraphics[width=4.54cm]{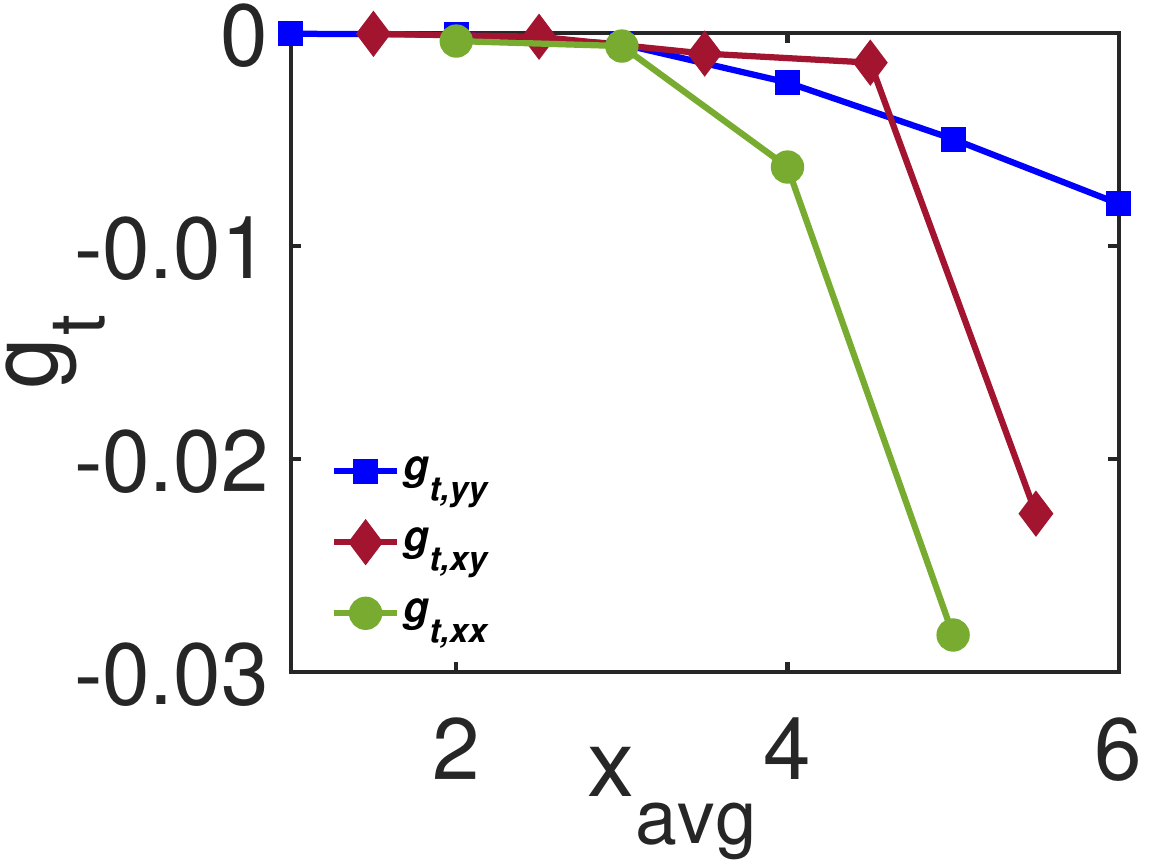}%
  \label{fig5b}%
}\qquad
\subfloat[]{%
  \includegraphics[width=4.54cm]{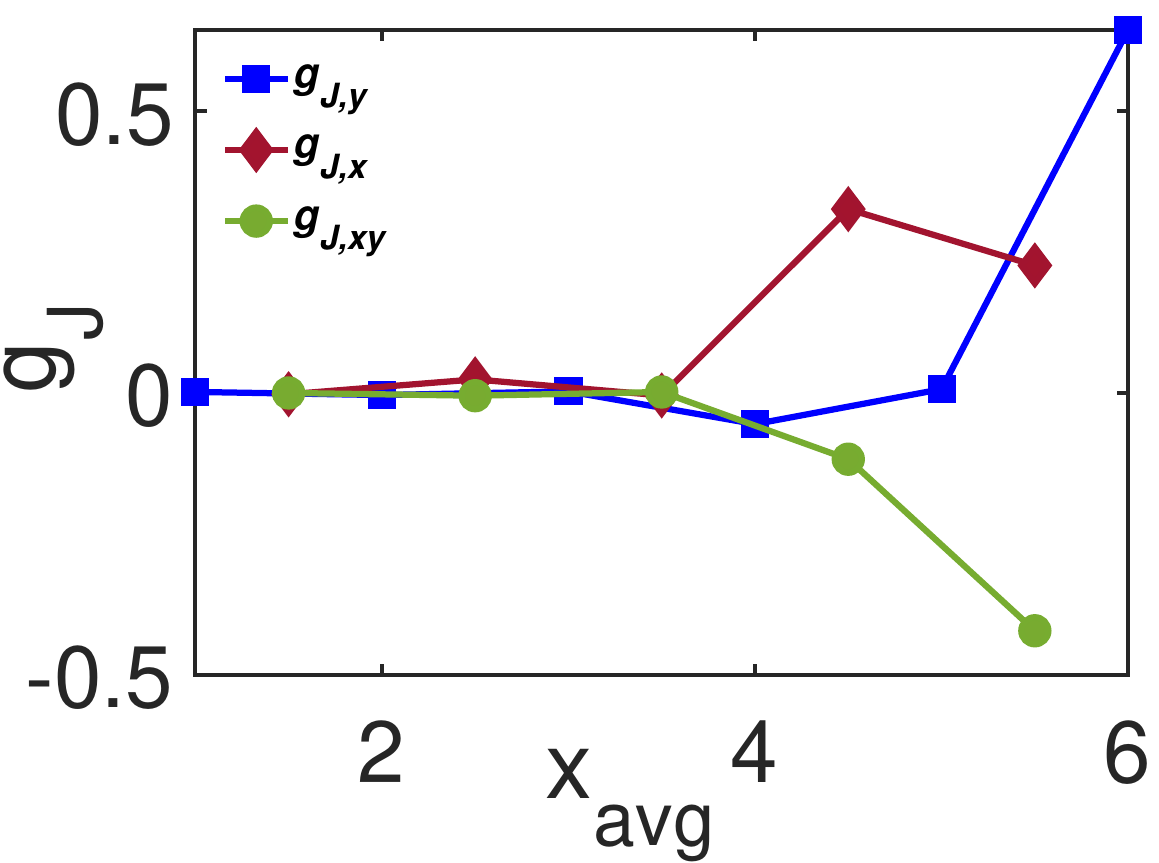}%
  \label{fig:evaluation:avgPrice}%
}
\subfloat[]{%
  \includegraphics[width=4.54cm]{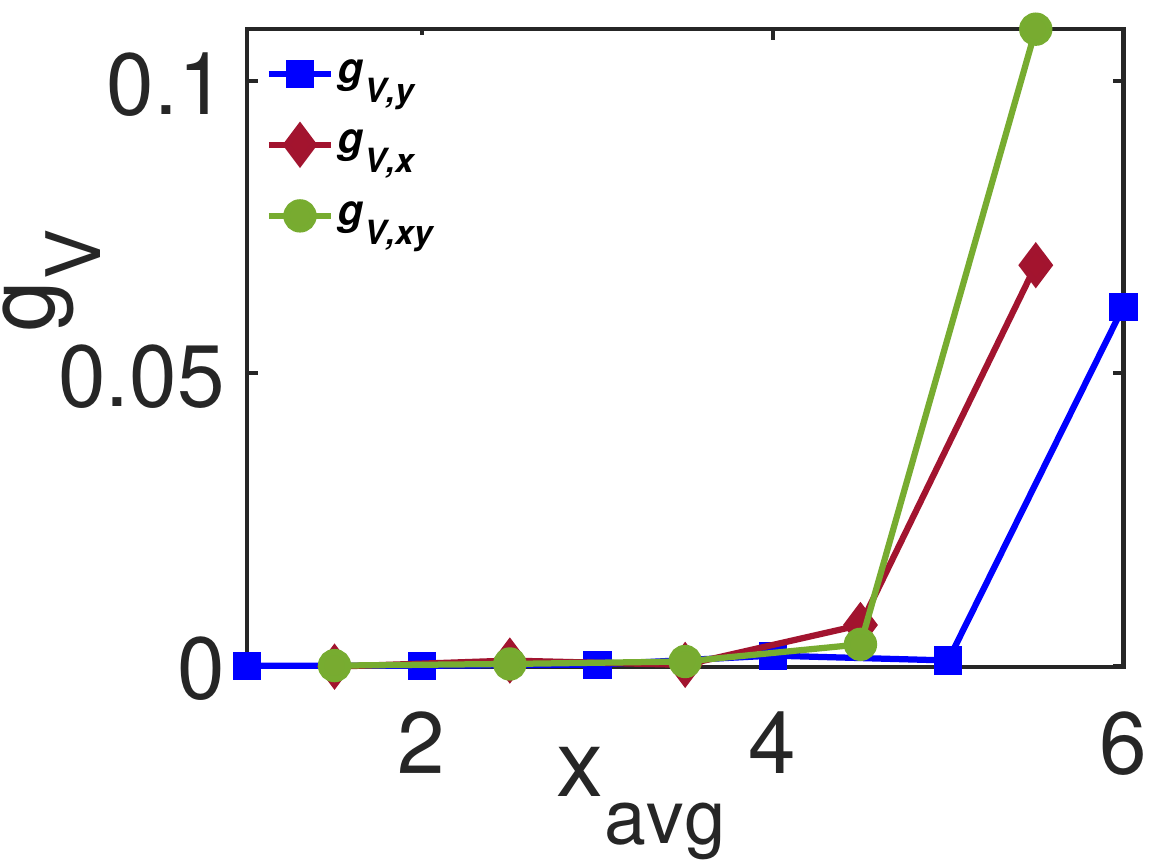}%
  \label{fig5b}%
}
\caption{\raggedright EH's couplings for the Hubbard model at half-filling ($U=4t_1$) for the subsystem geometry shown in Fig.~\ref{fig0}-d. LTA corrections are small (compared to the leading terms) everywhere, particularly away from $\partial A$. The particle-hole symmetry dictates the couplings in (b) to vanish. However, due to the finite truncation error of DMRG at $\chi = 2^{10}$, we obtain nonzero, though negligible values.}
\label{fig5}
\end{figure}
\begin{figure}[t]
\centering
\setlength{\belowcaptionskip}{-5pt}
\subfloat[]{%
  \includegraphics[width=4.54cm]{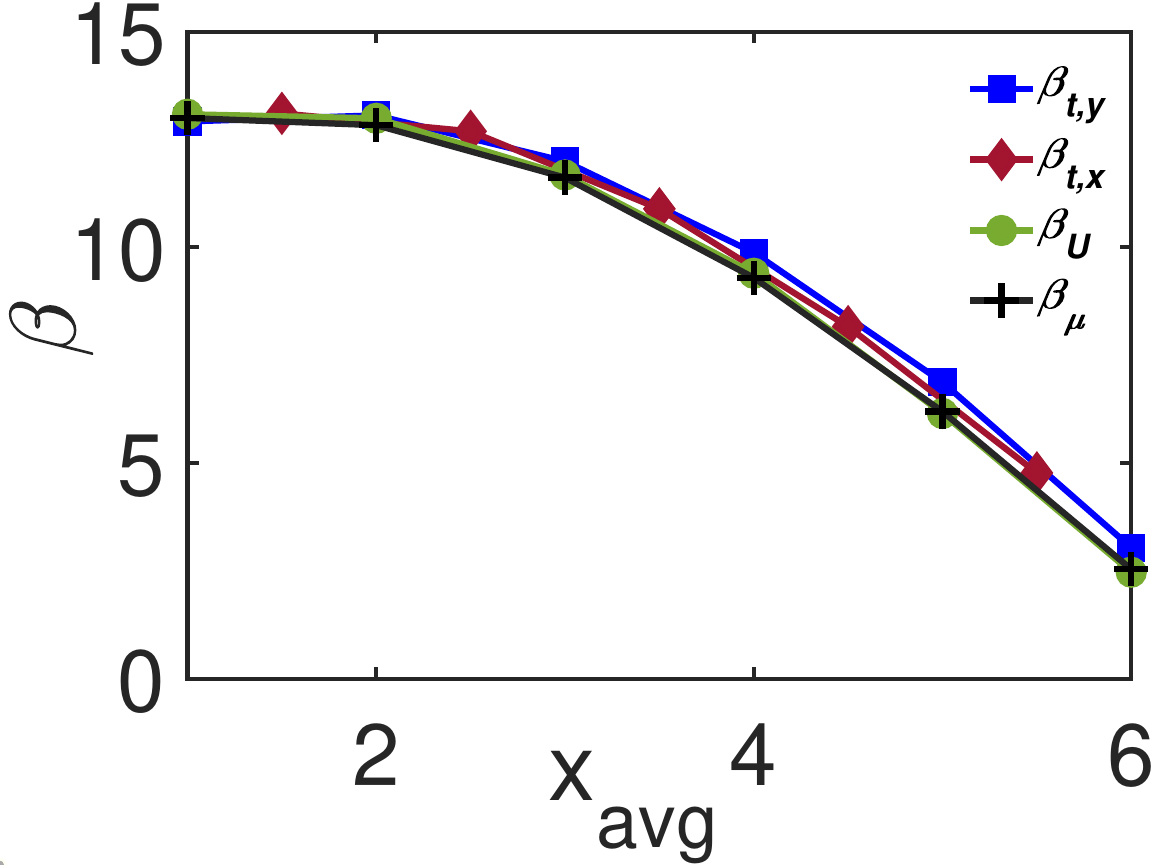}%
  \label{fig6a}%
}
\subfloat[]{%
  \includegraphics[width=4.54cm]{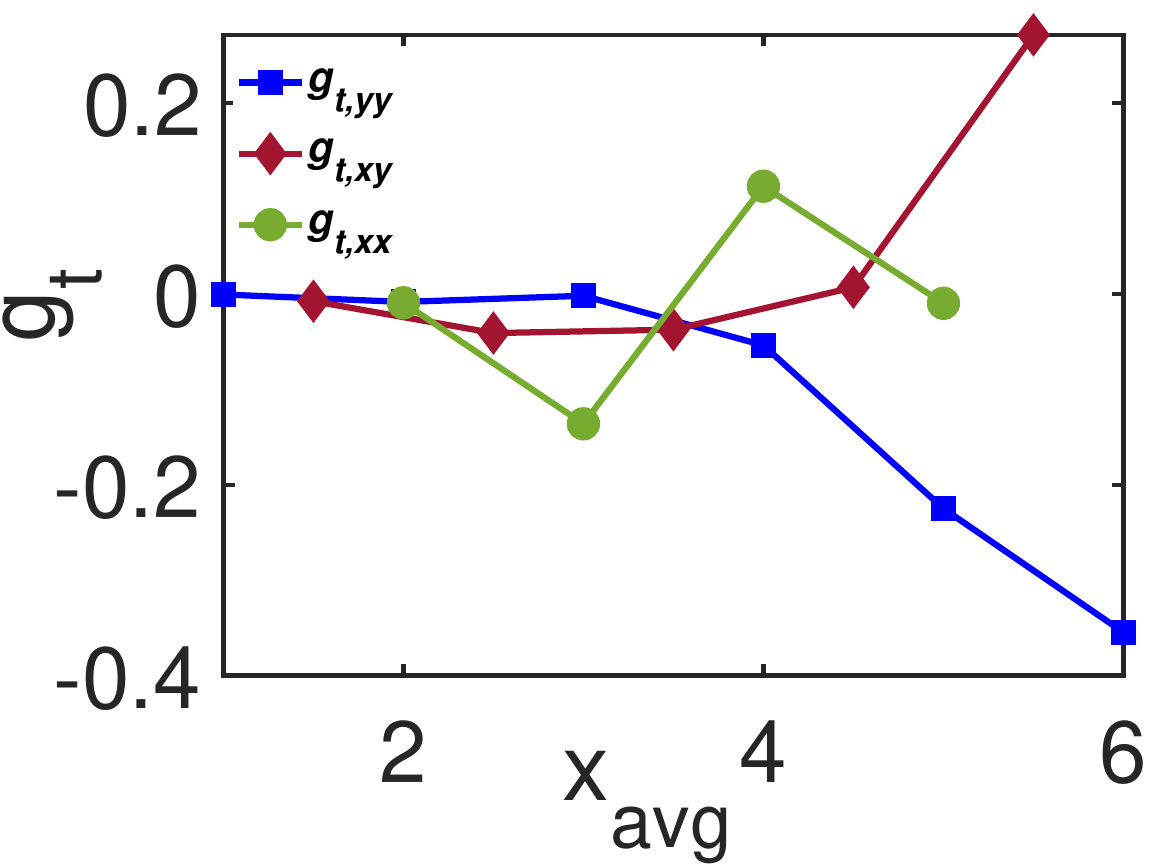}%
  \label{fig6b}%
}\qquad
\subfloat[]{%
  \includegraphics[width=4.54cm]{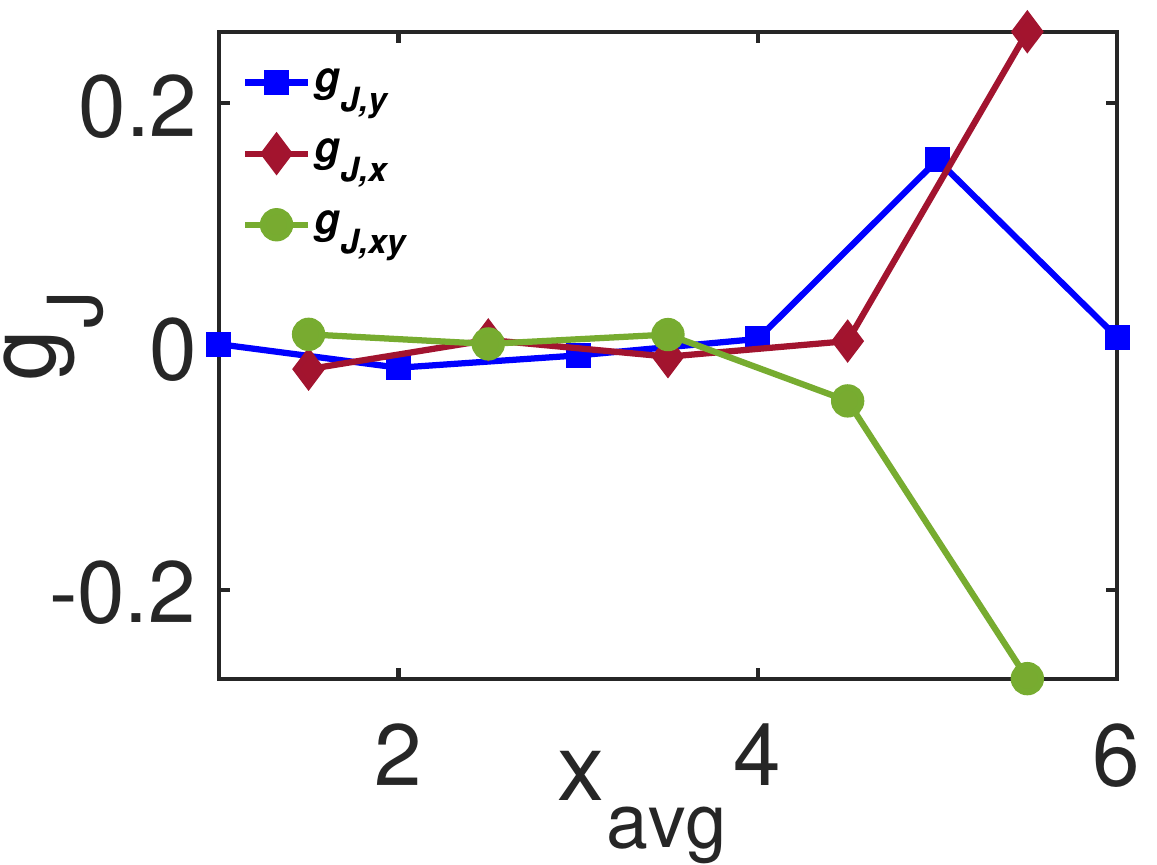}%
  \label{fig6c}%
}
\subfloat[]{%
  \includegraphics[width=4.54cm]{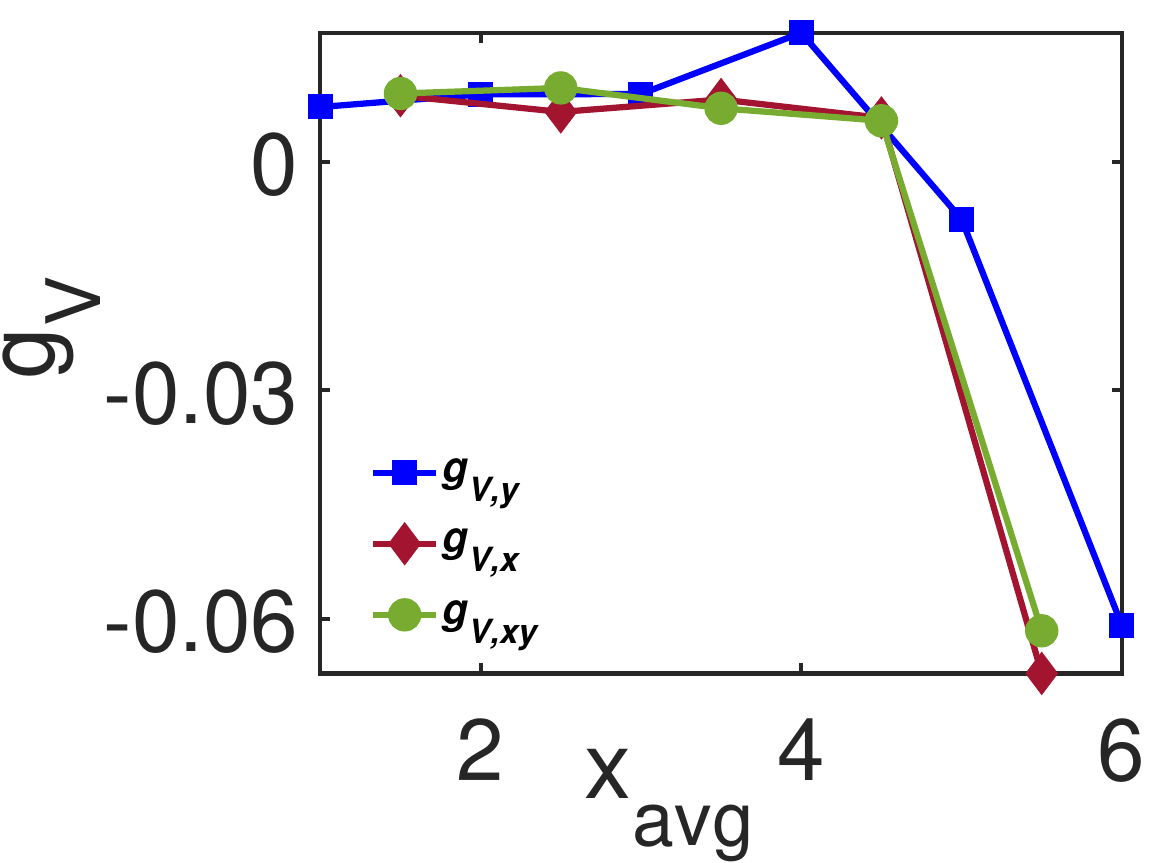}%
  \label{fig6d}%
}
\caption{\raggedright EH's couplings for the Hubbard model at $p=1/8$ doping level ($U=4t_1$, $\mu \approx -0.92t_1$) for the subsystem geometry shown in Fig.~\ref{fig0}-d.} 
\label{fig6}
\end{figure}
\bea
 K_A = &&-\sum_{{\bf i\neq j} \in A} g_{t,\bf ij}\, c_{\bf i,\sigma}^\dag c_{\bf j,\sigma} - \sum_{{\bf i} \in A, \sigma} g_{\mu,\bf i}\, n_{\bf i,\sigma}  \cr
&&+ \sum_{{\bf i} \in A} g_{U,\bf i} \para{n_{\bf i,\up}-\frac{1}{2}}\para{n_{\bf i,\dn}-\frac{1}{2}} \cr
&&+\sum_{{\bf i\neq j}\in A} \big(g_{V,\bf ij}\, \para{n_{\bf i}-1} \para{n_{\bf j}-1} + g_{J,\bf ij}\, {\bf S_{i} .  S_{j}}\big) + \cdots, ~~~~\label{EH2}
\eea
where $n_{\bf i} = n_{\bf i,\up}+n_{\bf i,\dn}$ denotes the total electron number on site $\bf i$, and $S^{x,y,z}_{\bf i} = \sum_{ab} \frac{1}{2}c_{{\bf i},a}^\dag \sigma^{x,y,z}_{a,b} c_{{\bf i},b}$ 
the three components of the spin operator at $\bf i$. Again, we have discarded higher order terms as we attain satisfactory results with the above structure. In the following, we consider both doped and undoped Hubbard models. In this section, we consider the geometry shown in Fig.~\ref{fig0}-d. 

\item Let us start with the half-filling case. The ground-state on the square lattice is described by a Néel anti-ferromagnetic spin order and the charge/Mott gap opens up at moderate values of $U$ \cite{varney2009quantum,vsimkovic2020extended}. 
Fig.~\ref{fig5}, summarizes our results for the optimum couplings of the EH. Here, motivated by LTA, we define the following inverse temperatures: $\beta_{t,x}\para{i_x+1/2} := g_{t,\bf i,i + \hat{x}}$, $\beta_{t,y}\para{i_x} := g_{t,\bf i,i + \hat{y}}$, and $\beta_{U}\para{i_x} := \frac{1}{U}g_{U,\bf i}$. According to Fig.~\ref{fig5}, they all fairly follow an identical curve, albeit by considering the previously discussed $1/2$ shift in the argument of $\beta_{t,x}$. In Figs.~\ref{fig5}b-d, we have shown the terms beyond LTA and again the locality of couplings is confirmed. Only near $\partial A$, the additional terms are non-negligible. Although, $g_{V,\bf ij}$ is insignificant everywhere, $g_{J,\bf ij}$ has decent values near $\partial A$, yet inferior to those of $g_{t,\bf \left<ij\right>}$ and $g_{U,\bf i}$. 

\item We now study the Hubbard model at $p=1/8$ doping which is expected to have a stripe order and some tendency towards superconductivity \cite{jiang2020ground,huang2017numerical,fradkin2015colloquium,zhou2017quantum}. The system is not expected to exhibit the Lorentz invariance or conformal symmetry for these symmetry breaking phases. At finite doping, we need to define $\beta_{\mu}\para{i_x} := \frac{1}{\mu}g_{\mu,\bf i}$ as well (for the current example: $\mu \approx -0.92t_1$). As Fig.~\ref{fig6} implies, various $\beta$'s follow the expected trend and the locality of couplings is again verified, albeit with growing corrections close to $\partial A$.

\end{itemize}

\noindent {\it Summary.}--- Our DMRG-based algorithm allowed us to access the second quantization form of the EH for several models and subsystem shapes. We showed that the EH is local and its dominant components are related to those of the Hamiltonian itself (more specifically the stress-energy tensor) up to a single smooth local (inverse) temperature profile and confirmed LTA. We studied the terms beyond LTA and demonstrated they are infinitesimal far away from $\partial A$ and relatively small near it. In the Appendix, we have provided more evidences which further corroborate our main findings. To our knowledge, the validity of LTA for the ground-state of local Hamiltonians for generic models that do not satisfy conformal algebra or even those with conformal symmetry but non-flat $\partial A$ is an unsolved problem despite active research. Our results suggest that LTA is perhaps a legitimate assumption and applicable to a broader class of problems. 

Our findings pave the way for several applications of LTA. For instance, it can be shown that LTA can practically solve the long-standing sign problem in quantum Monte Carlo and enable us to extract the ground-state properties of some unsolved interacting models. Furthermore, LTA can be employed to enhance the performance and increase the accuracy of the DMRG technique. It can also be used to recover the entire spectrum and eigenstates of an unknown Hamiltonian by having access to its reduced density matrix (or correlation functions) associated with a rather small subregion of that system~\cite{turkeshi2019entanglement}.

\noindent {\it Acknowledgements.}--- We gratefully acknowledge helpful discussions with M. Dalmonte, A. Lucas, S. Nezami, Z. Nussinov, H. Yarloo, A. Shahbazi, E. Huang, M. Kargarian, A. Rezakhani and S. Alipour. AV acknowledges the Gordon and Betty Moore Foundation’s EPiQS Initiative through Grant GBMF4302 and Stanford Center for Topological Quantum Physics for partial financial support and hospitality during the completion of this work. MSV acknowledges the financial support from Pasargad Institute for Advanced Innovative Solutions (PIAIS) under supporting Grant scheme (Project No. SG1-RCM2001-01).

\subsection{APPENDIX}

In this appendix, we will delve into the details of our algorithm and discuss the advantages and disadvantages of a number of cost function candidates along with their pairwise comparison, and present more results on the entanglement Hamiltonian (EH).

\subsection{A. Entanglement Hamiltonian in the truncated Hilbert space}
\label{SM_A}

In general, the entanglement Hamiltonian (EH) associated with subsystem $A$ which is defined as $K_A := -\log \rho_A$, can be expanded in terms of a complete basis of operators (not necessarily local) as follows: (the tensor product of Pauli matrices, $\sigma_{a}$, $a=x,y,z$ with $\sigma_0 = \mathbb{1}$, can generate a basis for all possible operators):

\bea
K_A = \sum_{\alpha} g_{\alpha} \hat{O}_{\alpha}
\eea

If we are given the reduced density matrix, $\rho_A$, we can compute its logarithm (which requires a lot of considerations and special care when performed numerically) and achieve $K_A$ (up to computer's round-off error). Having $K_A$ available, we can easily find the expansion coefficients, $g_{\alpha}$, via the following relation:

\bea
&& \vec{g} =  M^{-1}\vec{J} 
\eea
where
\bea
&& J_{\beta} = {\rm Tr}_A \para{K_A \hat{O}_{\beta}^\dag},\cr
&& M_{\alpha \beta} = {\rm Tr}_A \para{\hat{O}_{\alpha} \hat{O}_{\beta}^\dag}
\eea

In the exact diagonalization (ED) method $M_{\alpha,\beta} \propto \delta_{\alpha,\beta}$. Therefore, $g_{\alpha} \propto {\rm Tr}_A \para{K_A \hat{O}_{\alpha}^\dag}$. Consequently, we do not need to consider other operators if we are interested in reading the coefficient for a specific $\alpha$. 

On the other hand, in the density-matrix-renormalization-group (DMRG) algorithm, instead of $\rho_A$, and $\hat{O}_{\alpha}$, we have to deal with $\overline{\rho_A}$, and $\overline{\hat{O}_A}$ which are their counterparts in the truncated Hilbert space and are defined as:
\bea
\overline{\rho_A} = T_A^\dag \rho_A T_A,
\eea 
and similarly for other operators, where $T_A$ denotes the truncation (a.k.a. projection) operator. As before, we define $\overline{K_A} = - \log \overline{\rho_A}$. Because of the numerous truncations involved in DMRG, the situation is now more complicated for a few reasons: (i) The matrix $M$ is not diagonal, neither sparse. Due to consecutive truncations inherent to the DMRG method, most of operators have non-vanishing overlaps. Therefore, we must consider all possible operators, including highly non-local ones such as string or brane operators. (ii) $M$ can be singular and have zero eigenvalues. As a result, it might not be invertible. (iii) The above method applied to DMRG is very sensitive to various sources of numerical noises and errors, such as the truncation, as well as the round-off error. 

Besides the possibility of singular $M$, another main difficulty of applying the above algorithm to DMRG is the annoying part which requires taking all possible operators into consideration. Below, we easily demonstrate that if $K_A$ contains a few relevant and dominant terms, then $\overline{K_A}$ contains exactly the same couplings and structure. To this end, recall that $T_A$ is achieved upon concatenating the dominant eigenvectors of  $\rho_A$. Therefore,

\bea
\overline{\rho_A} = T_A^\dag \rho_A T_A = T_A^\dag e^{-K_A} T_A = e^{- T_A^\dag K_A T_A}.
\eea

Therefore,
\bea
\overline{K_A} := -\log \overline{\rho_A} = T_A^\dag K_A T_A.
\eea
Accordingly,
\bea
\overline{K_A} = \sum_{\alpha} g_{\alpha}T_A^\dag \hat{O}_{\alpha}T_A = \sum_{\alpha} g_{\alpha} \overline{\hat{O}_{\alpha}},
\eea
hence, assuming $\overline{K_A} = \sum_{\alpha} \overline{g_{\alpha}}\overline{\hat{O}_{\alpha}}$ :
\bea
\overline{g_{\alpha}} = g_{\alpha}.
\eea

\subsection{B. Algorithm and cost function}
\label{SM_B}

In the above mentioned method, for ED, we can ignore insignificant couplings since $M$ is diagonal. Nevertheless, when we apply this method to DMRG, we have to retain all terms, no matter how infinitesimal they are due to the complex form of $M$. Therefore, we must come up with a better algorithm to find $g_{\alpha}$ without having to consider all irrelevant terms. For that purpose, we must consider a valid cost function. From our physical intuitions and expectations, we can think of the following three choices (as of now, we drop the overline sign and keep in mind that all operators are defined in the truncated Hilbert space):

\begin{itemize}

\item Hilbert-Schmidt distance between the Green's functions (GFs): $\Delta_1 = {\rm Tr}\para{G_A-\widetilde{G_A}}^2$, where $G_{A}(\alpha,\beta) = {\rm Tr}_A \para{O_{\alpha}^\dag O_{\beta} \rho_A}$, and $\widetilde{G_{A}}(\alpha,\beta) = {\rm Tr}_A \para{O_{\alpha}^\dag O_{\beta} \widetilde{\rho_A}}$. Here $\rho_A$ denotes the reduced density matrix (RDM) achieved via DMRG for the desired subsystem and $\widetilde{\rho_A}$ denotes the one by combining the basis operators ($O_{\alpha}$) with $g_{\alpha}$ coefficients that are yet to determine. The basis of this method is that the RDM contains all the information about the equal time correlation functions within the subsystem. Thus, if we find a RDM which recovers all the correlation functions correctly, it must be identical to the actual one. 

We would like to emphasize that in DMRG, due to finite truncation error, the RDM yields more reliable results for the expectation value of simple operators (e.g., two-point correlation functions for short and intermediate distances) and becomes less reliable for more complex operators or at long distances. Therefore, to avoid overfitting to numerical errors, instead of considering all basis operators in the evaluation of $\Delta_1$, we only consider the most physically relevant operators, i.e., simple operators motivated by symmetry considerations, etc. For example, for the Hubbard model, we consider the following components first: 
\bea
&&G_{A,\rm t}\para{i,j} = \sum_{\sigma} \left<c_{i,\sigma}^\dag c_{j,\sigma}\right>,\cr
&&G_{A,\rm \mu}\para{i} = \left<n_{i} \right>,\cr
&& G_{A,\rm U}\para{i} = \left<n_{i,\up} n_{i,\dn}\right>,\cr
&& G_{A,\rm J}\para{i,j} = \left<{\bf S}_{i} . {\bf S}_{j}\right>,\cr
&&G_{A,\rm V}\para{i,j} = \left<n_{i} n_{j}\right>,\nonumber
\eea
and similarly for $\tilde{G}$. We then evaluate $\eta_{a} = {\rm Tr}\para{G_{A,a}-\widetilde{G_{A,a}}}^2$ and by combining them:

\bea
&& \Delta_1 = w_t \eta_t + w_{\mu} \eta_{\mu} + w_U \eta_U + w_J \eta_J + w_V \eta_V.
\eea
The exact values of $w_t, w_{\mu}, \cdots$ are not crucial as long as they all have the same order of magnitude. Nonetheless, in most computations, we choose $w_t = w_U =w_{\mu} = w_{J} = w_V = 1$ for the weights. 

\item Quantum relative entropy (QRE) of the two reduced density matrices: $\Delta_2 = {\rm Tr}\para{\rho_A \log \rho_A - \rho_A \log \widetilde{\rho}_A}$. In this method (which is closely related to the next cost function), we try to tune the couplings such that $\widetilde{\rho_A}$'s matrix form becomes very close to $\rho_A$'s. This cost function converges significantly fast, in both ED and DMRG method. In ED where we do not have to deal with truncation errors, QRE is the superior cost function and achieves correct results. However, for DMRG, (like $\Delta_3$ below), it also suffers from overfitting to numerical errors, i.e., those parts of $\rho_A$ which will change upon increasing the bond dimension of DMRG, $\chi$ (i.e., the number of retained basis states of the Hilbert space). It is these matrix elements which are responsible for the issues related to the expectation value or $n-$point correlation functions of complex operators explained above. When $\chi$ is large enough (e.g., when the truncation error becomes less than $10^{-10}$), it yields results consistent with $\Delta_1$'s.

\begin{figure}[t]
\centering
\includegraphics[width=5.0cm]{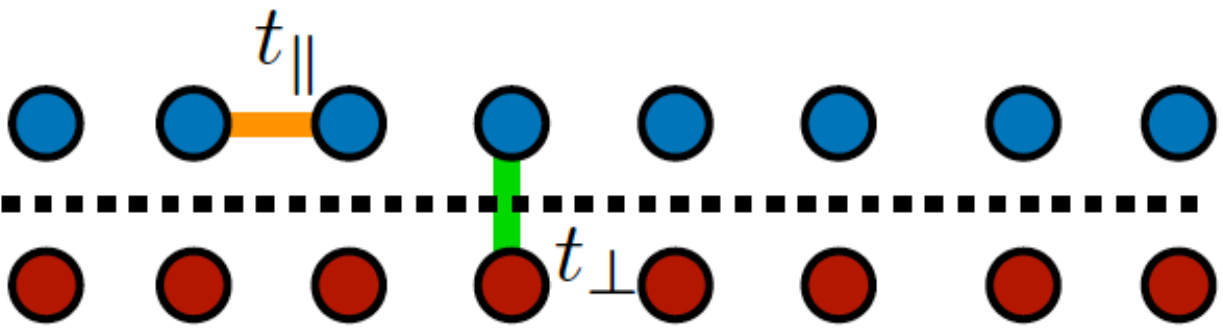}%
\caption{\raggedright We study the Hubbard model on this ladder for $U=4$, $t_{\perp} = 0.5$, and $t_{\parallel} = 1$ at half filling ($\mu =0$). Subsystem $A$, whose EH is desired, is denoted by blue sites.} 
\label{fig_SM_0a}
\end{figure}

\begin{figure}[t]
\centering
\subfloat[]{%
  \includegraphics[width=4.0cm]{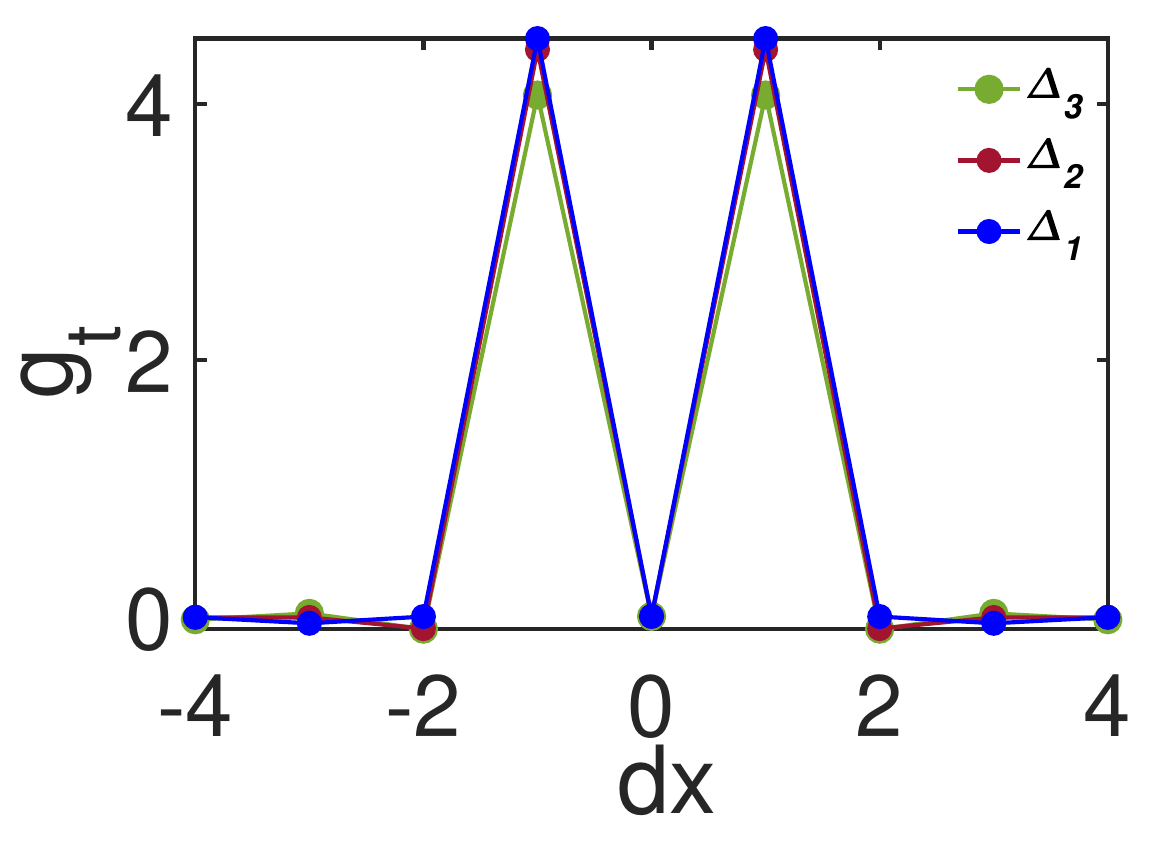}%
  \label{fig_SM_1a_a}%
}
\subfloat[]{%
  \includegraphics[width=4.0cm]{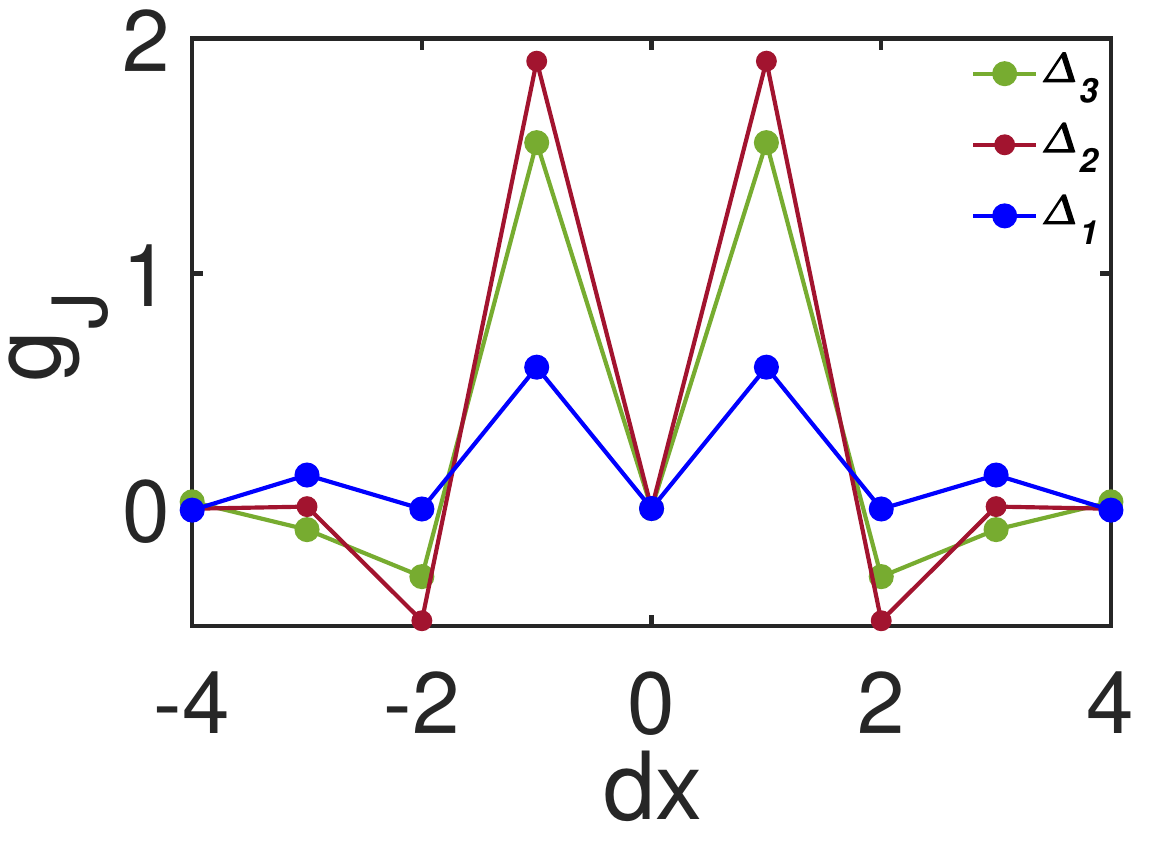}%
  \label{fig_SM_1a_b}%
}\qquad
\subfloat[]{%
  \includegraphics[width=4.0cm]{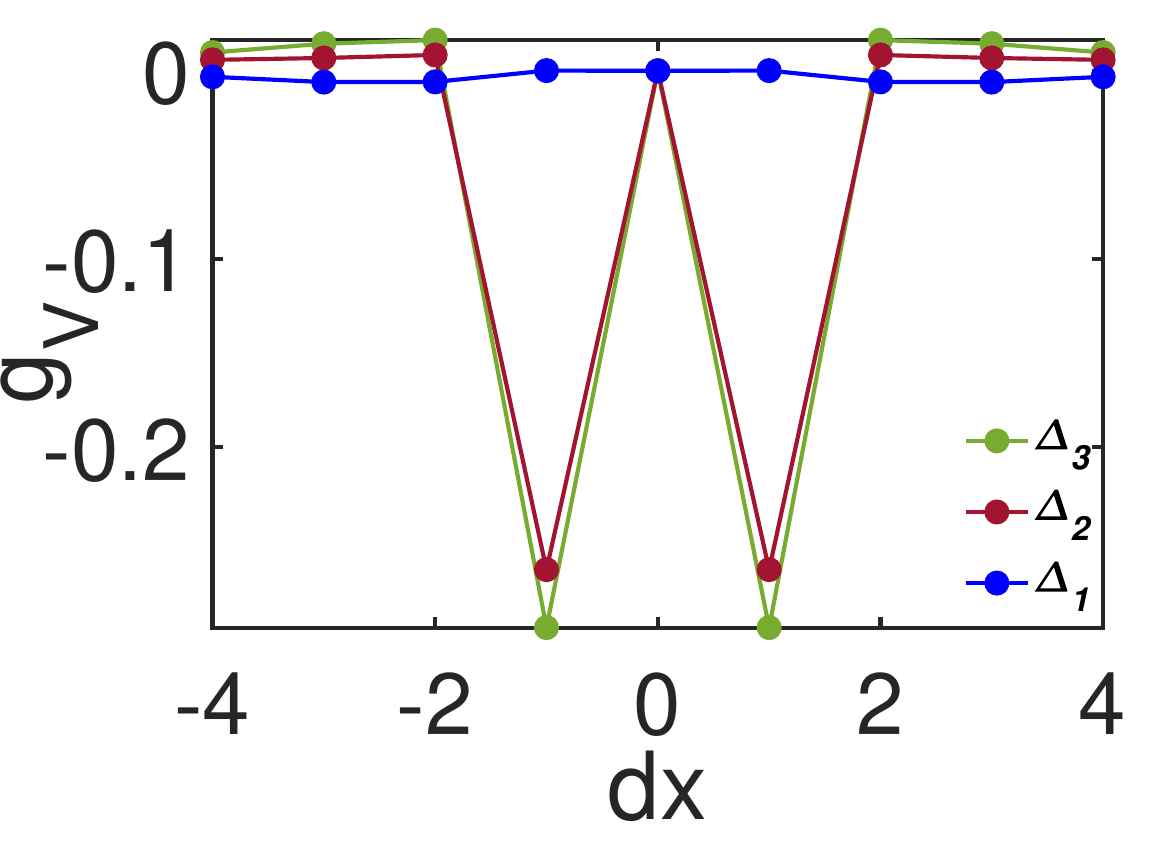}%
  \label{fig_SM_1a_c}%
}
\subfloat[]{%
  \includegraphics[width=4.0cm]{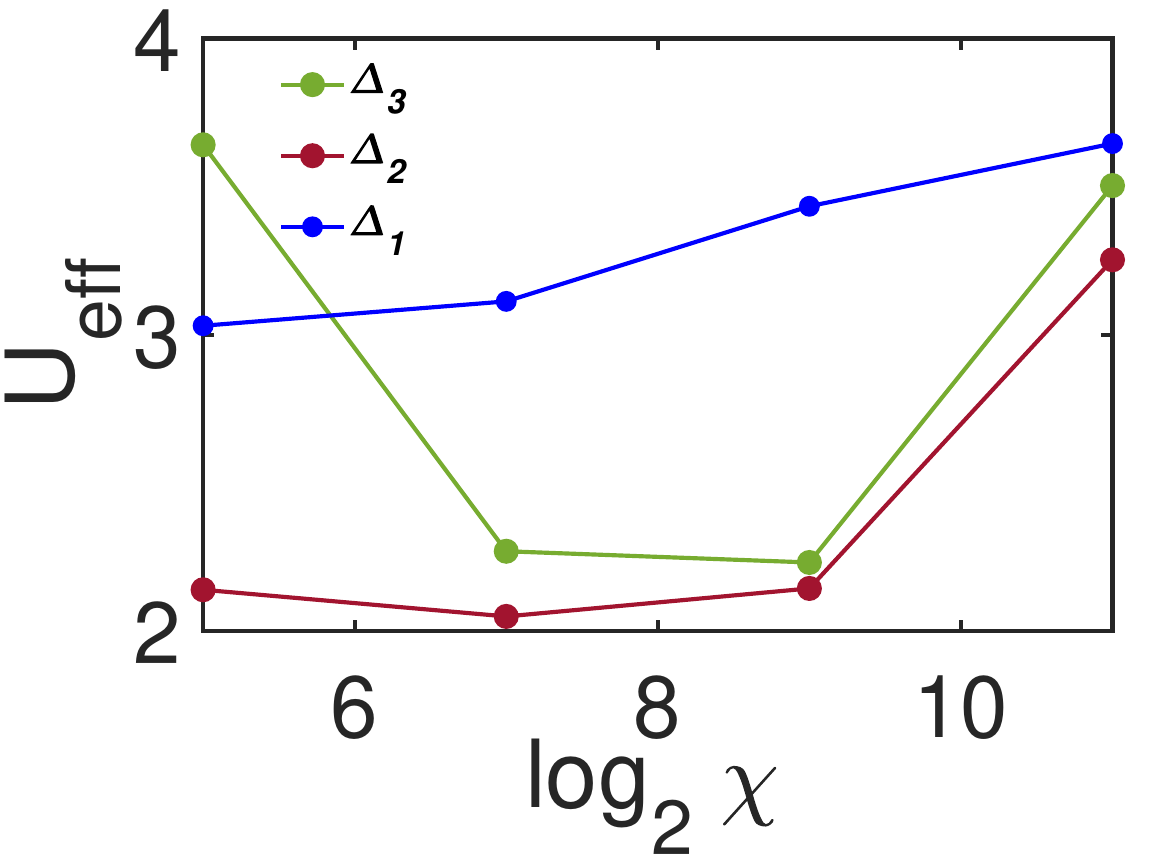}%
  \label{fig_SM_1a_d}%
}
\caption{\raggedright $K_A$'s couplings, for the geometry illustrated in Fig.~\ref{fig_SM_0a}, obtained via applying different cost functions and for various bond dimensions. Couplings are translational invariant due to the subsystem geometry. (a-c) $g_{t,dx}$, $g_{J,dx}$, and $g_{V,dx}$ achieved via GF distance ($\Delta_1$), QRE distance ($\Delta_2$), and RDM distance ($\Delta_3$) for $\chi=2^{11}$. Also, the corresponding renormalized onsite couplings, $g_{U,0}$ are $16.4, 14.4, 14.3
$ for $\Delta_1$, $\Delta_2$ and $\Delta_3$, respectively. See the text for their definitions. (d) The variation of $U_{\rm eff} := \frac{g_{t,1}}{g_{U,0}}$ versus $\log_2\chi$. As this plot clearly indicates, $\Delta_1$ is the most reliable cost function for DMRG and exhibits least variations.} 
\label{fig_SM_1a}
\end{figure}

\item Hilbert-Schmidt distance between the two reduced density matrices: $\Delta_3 = {\rm Tr}\para{\rho_A-\widetilde{\rho_A}}^2$. Similar to $\Delta_2$, in this method we try to tune the couplings such that $\widetilde{\rho_A}$'s matrix form becomes very close to $\rho_A$'s. This cost function is slowly converging even for the ED where no truncation is involved. Moreover, for DMRG, similar to $\Delta_2$, it suffers from overfitting to numerical errors and its results are sensitive to the bond dimension, especially for small values of $\chi$.

\end{itemize}

In section D of this Appendix we compare the results achieved via all three cost functions for the two-leg ladder Heisenberg and Hubbard models for several bond dimensions. Our results suggest that for large bond dimensions, all three methods yield consistent outcomes. However, for relatively small bond dimensions, it is $\Delta_1$ which performs better and results in couplings which are more consistent with the results of larger bond dimensions.

\begin{figure}[t]
\centering
\includegraphics[width=8.0cm]{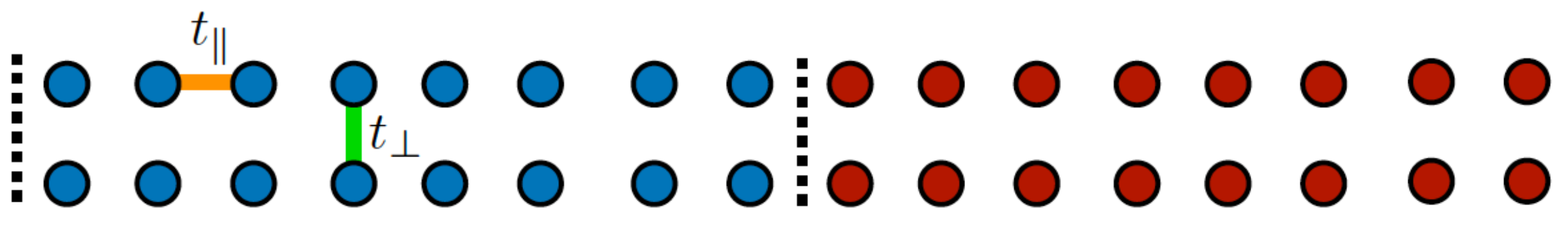}%
\caption{\raggedright We study the Hubbard model on this ladder for $U=4$, $t_{\perp} = 2$, and $t_{\parallel} = 1$ at half filling ($\mu =0$). Subsystem $A$, whose EH is desired, is denoted by blue sites.} 
\label{fig_SM_0b}
\end{figure}

\begin{figure}[t]
\centering
\subfloat[]{%
  \includegraphics[width=4.0cm]{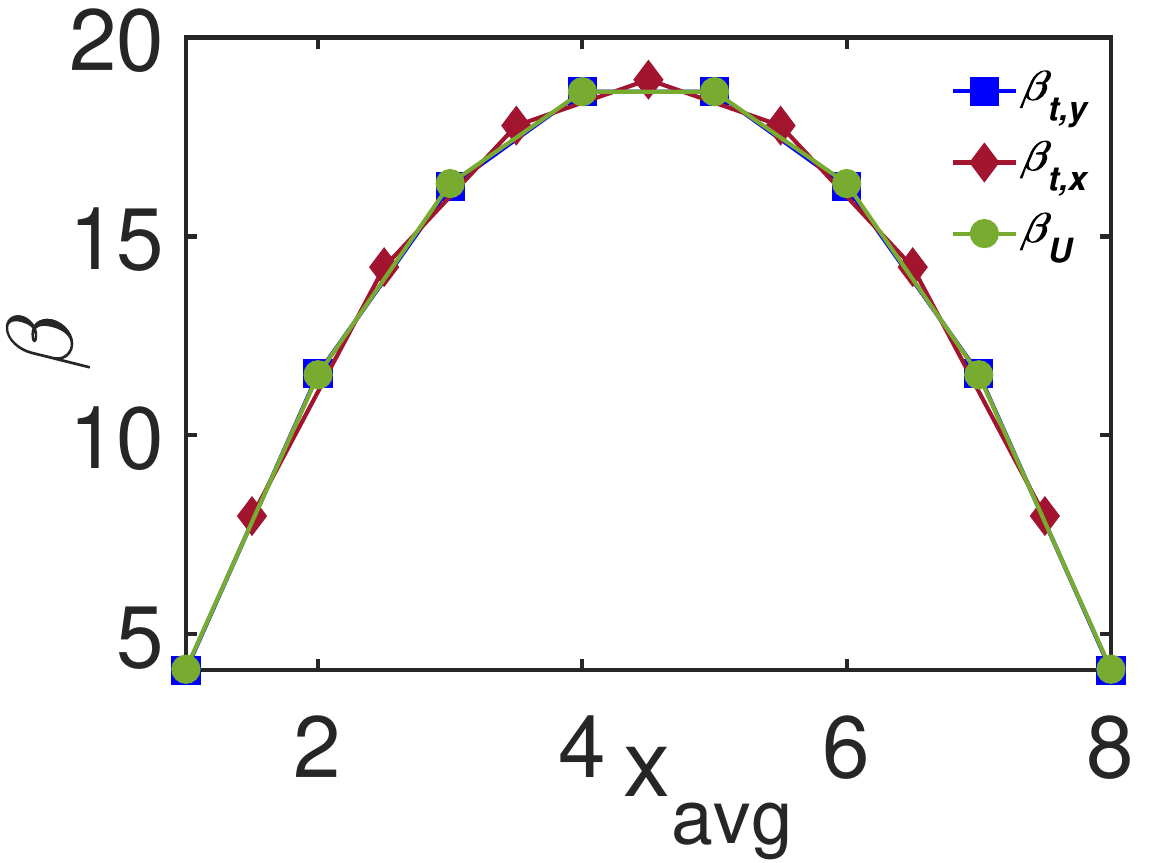}%
  \label{fig_SM_1b_GF_a}%
}
\subfloat[]{%
  \includegraphics[width=4.0cm]{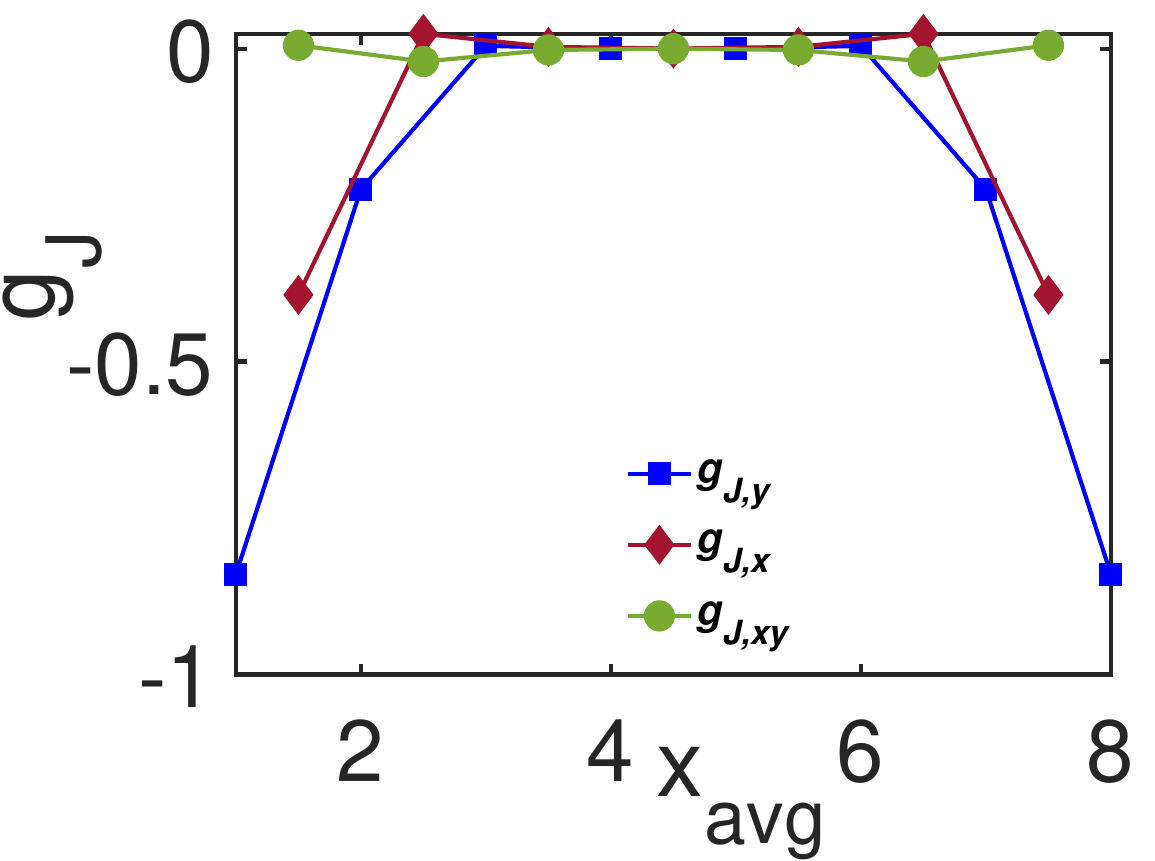}%
  \label{fig_SM_1b_GF_b}%
}\qquad
\subfloat[]{%
  \includegraphics[width=4.0cm]{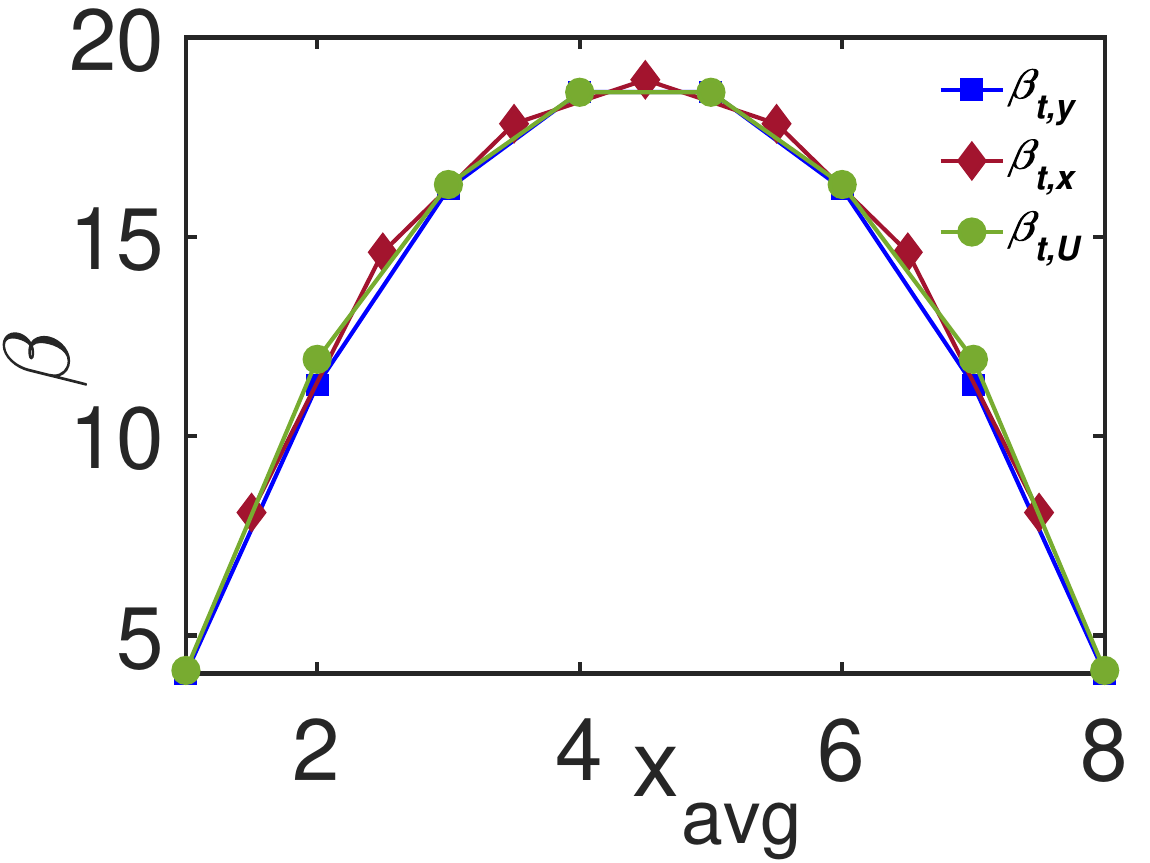}%
  \label{fig_SM_1b_GF_c}%
}
\subfloat[]{%
  \includegraphics[width=4.0cm]{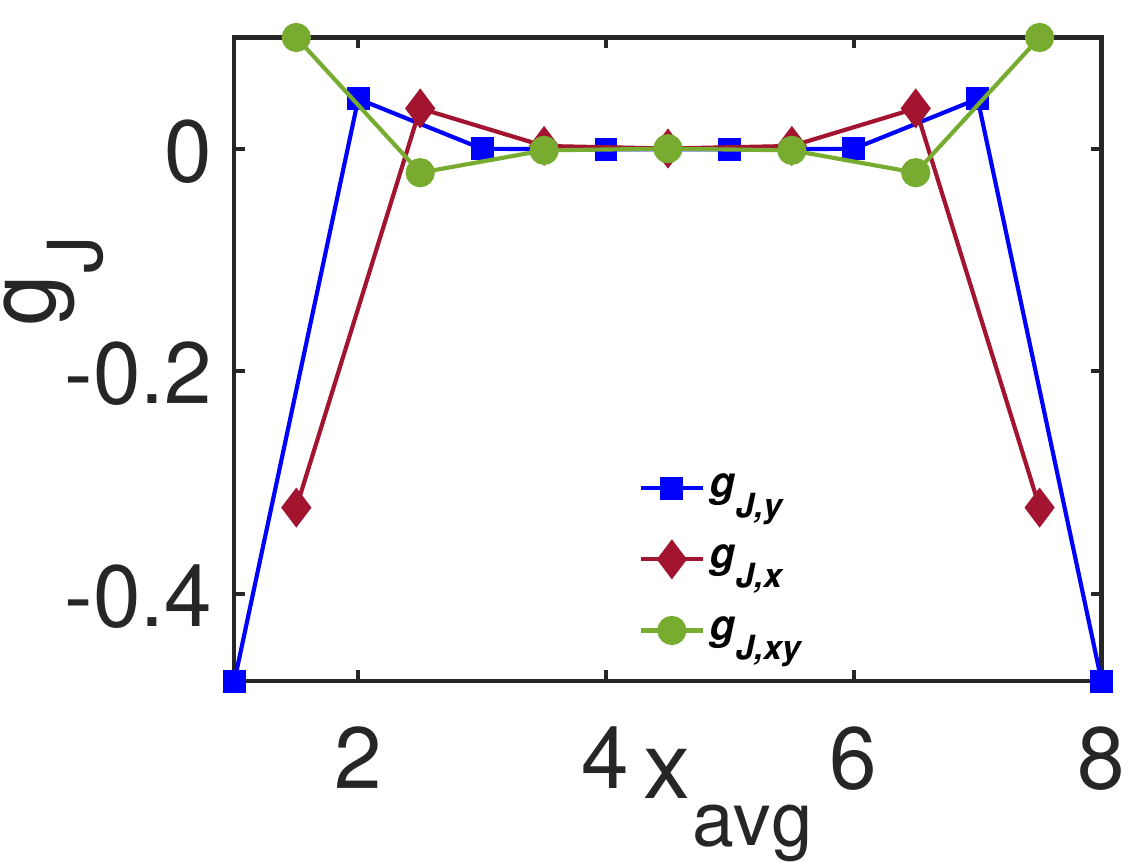}%
  \label{fig_SM_1b_GF_d}%
}
\caption{\raggedright $K_A$'s couplings, for the geometry illustrated in Fig.~\ref{fig_SM_0b}, obtained via applying GF distance ($\Delta_1$ cost function) and for $\chi = 2^{11}$, and $\chi = 2^9$. (a) Various $\beta$ profiles for $\Delta_1$ cost function for $\chi = 2^{11}$ (see the main text for their definitions). (b) Most significant corrections to LTA corrections, $g_{J}$ for $\Delta_1$ cost function for $\chi = 2^{11}$. The second and third neighbors' corrections to $g_{t}$ are negligible due to the particle-hole symmetry. Moreover, we found $g_V$ to be irrelevant as well, and that is why they are absent in this and the following two figures. (c-d) Same as (a-b) but for $\chi = 2^9$.} 
\label{fig_SM_1b_GF}
\end{figure}

\begin{figure}[t]
\centering
\subfloat[]{%
  \includegraphics[width=4.0cm]{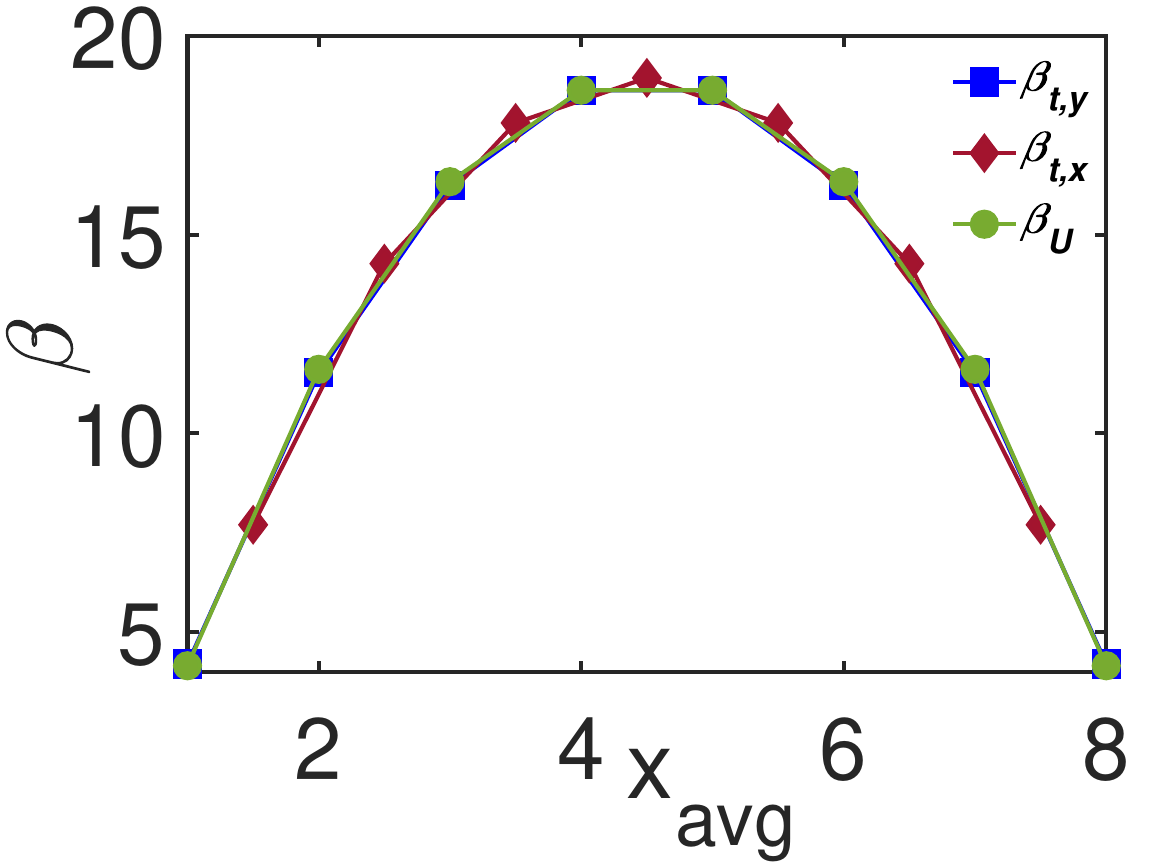}%
  \label{fig_SM_1b_RE_a}%
}
\subfloat[]{%
  \includegraphics[width=4.0cm]{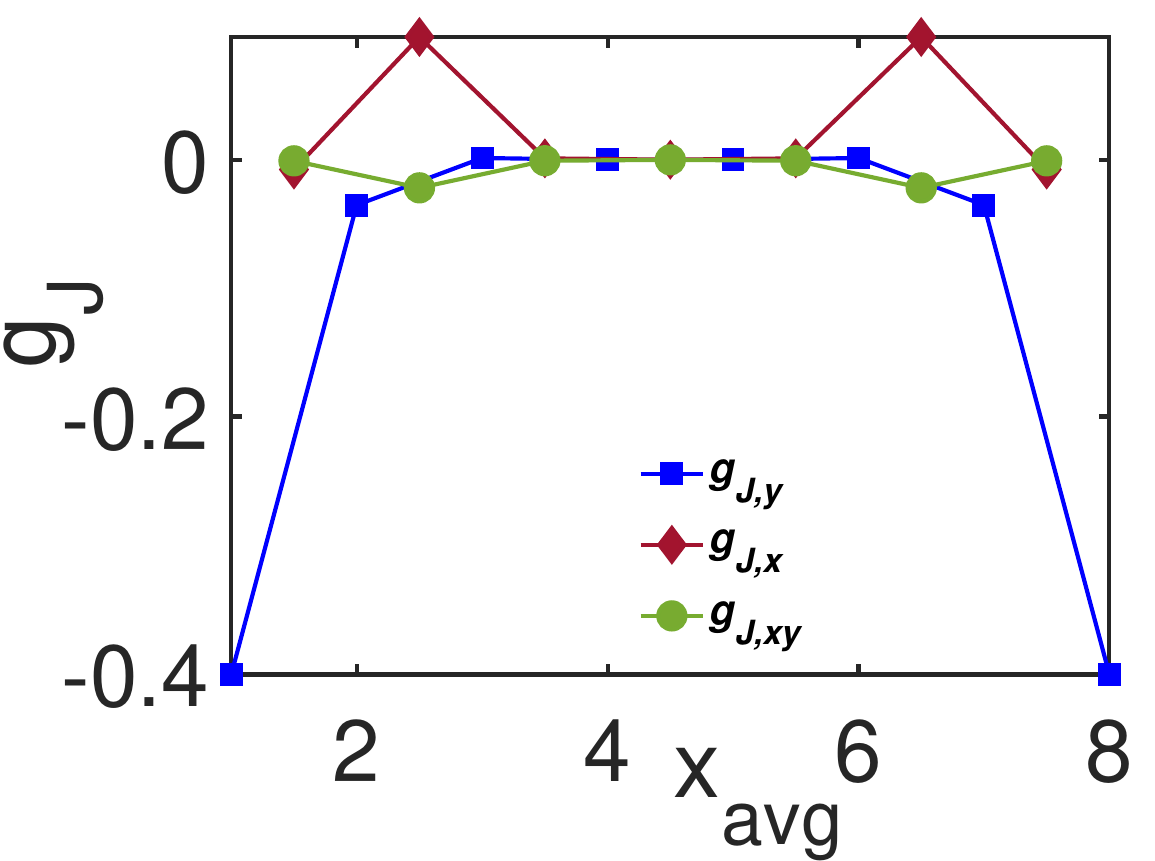}%
  \label{fig_SM_1b_RE_b}%
}\qquad
\subfloat[]{%
  \includegraphics[width=4.0cm]{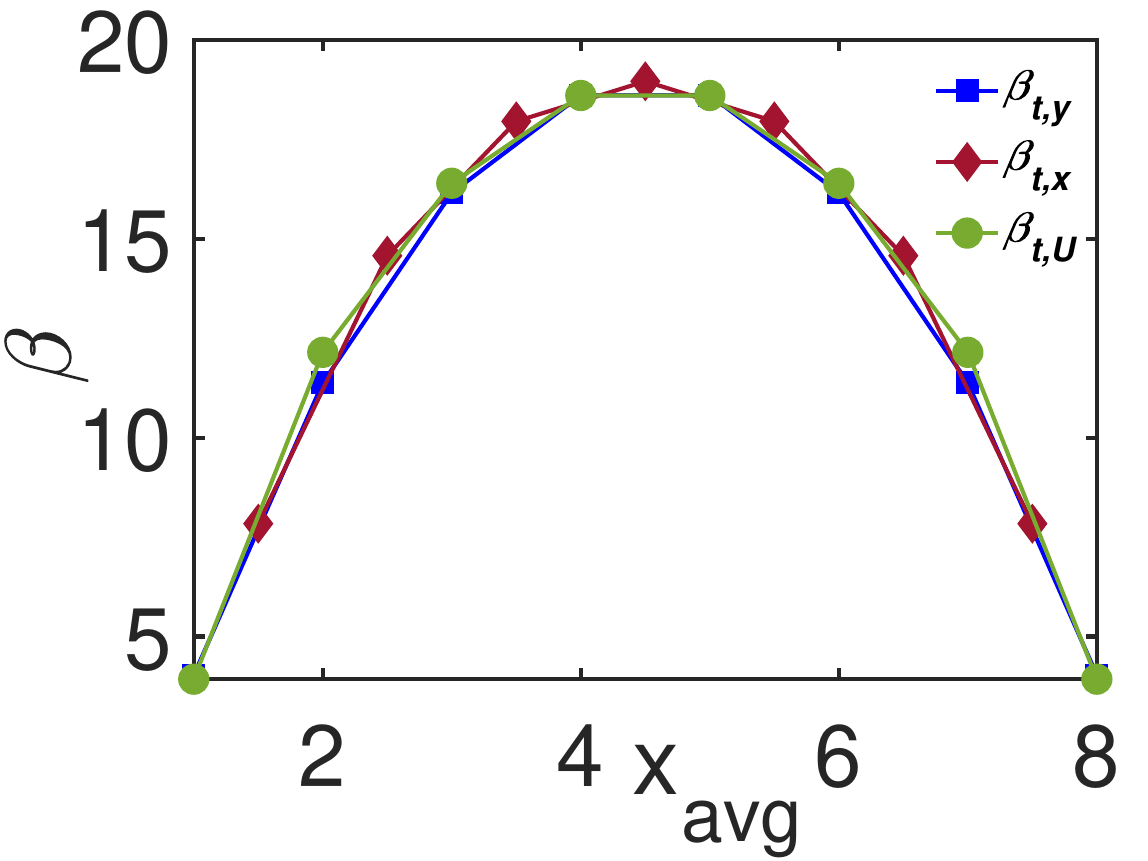}%
  \label{fig_SM_1b_RE_c}%
}
\subfloat[]{%
  \includegraphics[width=4.0cm]{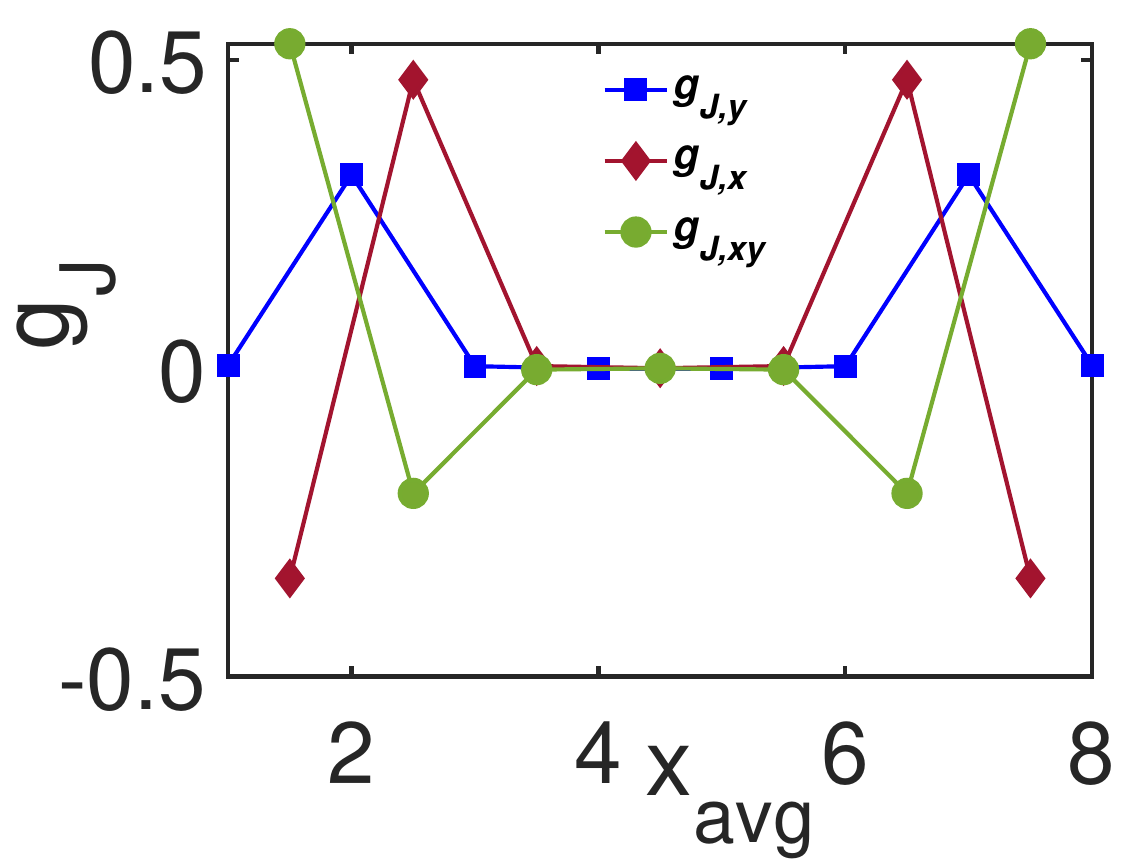}%
  \label{fig_SM_1b_RE_d}%
}
\caption{\raggedright Same as Fig.~\ref{fig_SM_1b_GF} but for $\Delta_2$ cost function (QRE). Likewise, (a) and (b) are achieved by considering $\chi=2^{11}$, while (c) and (d) by $\chi=2^9$} 
\label{fig_SM_1b_RE}
\end{figure}

\begin{figure}[t]
\centering
\subfloat[]{%
  \includegraphics[width=4.0cm]{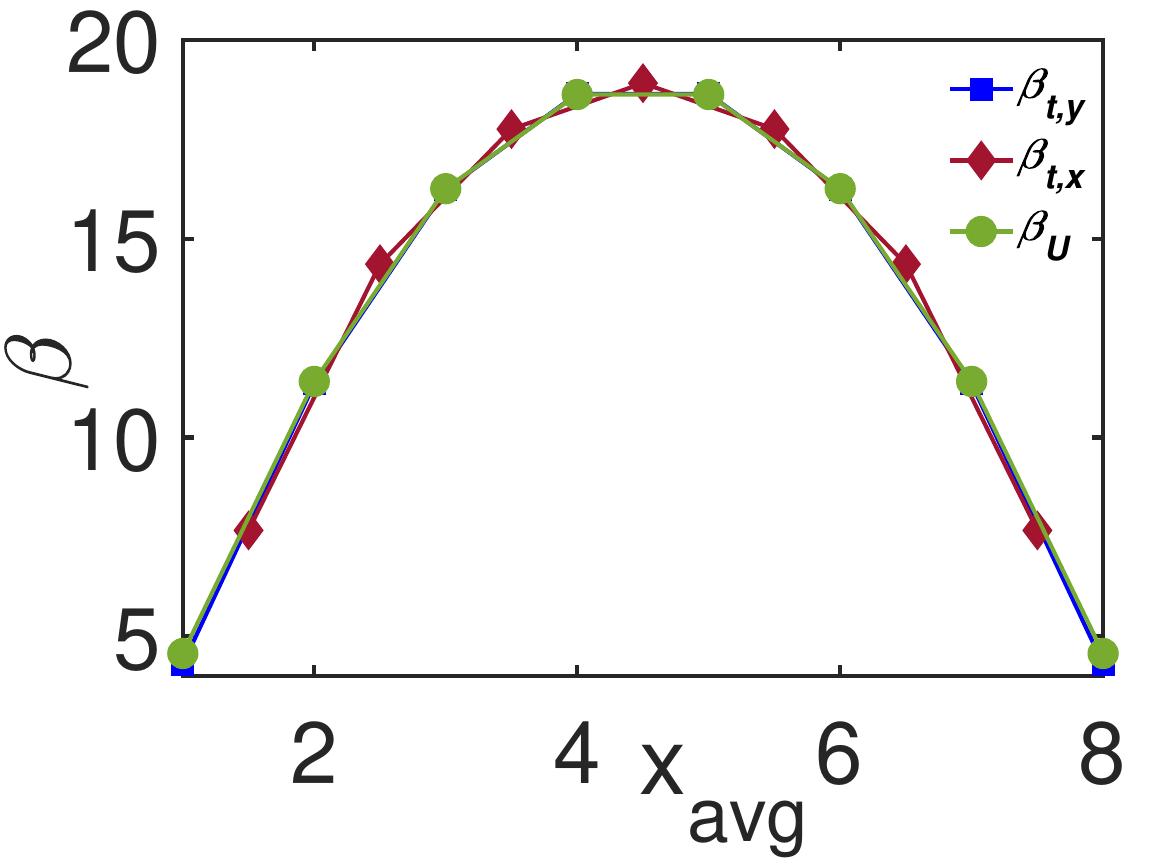}%
  \label{fig_SM_1b_rho_a}%
}
\subfloat[]{%
  \includegraphics[width=4.0cm]{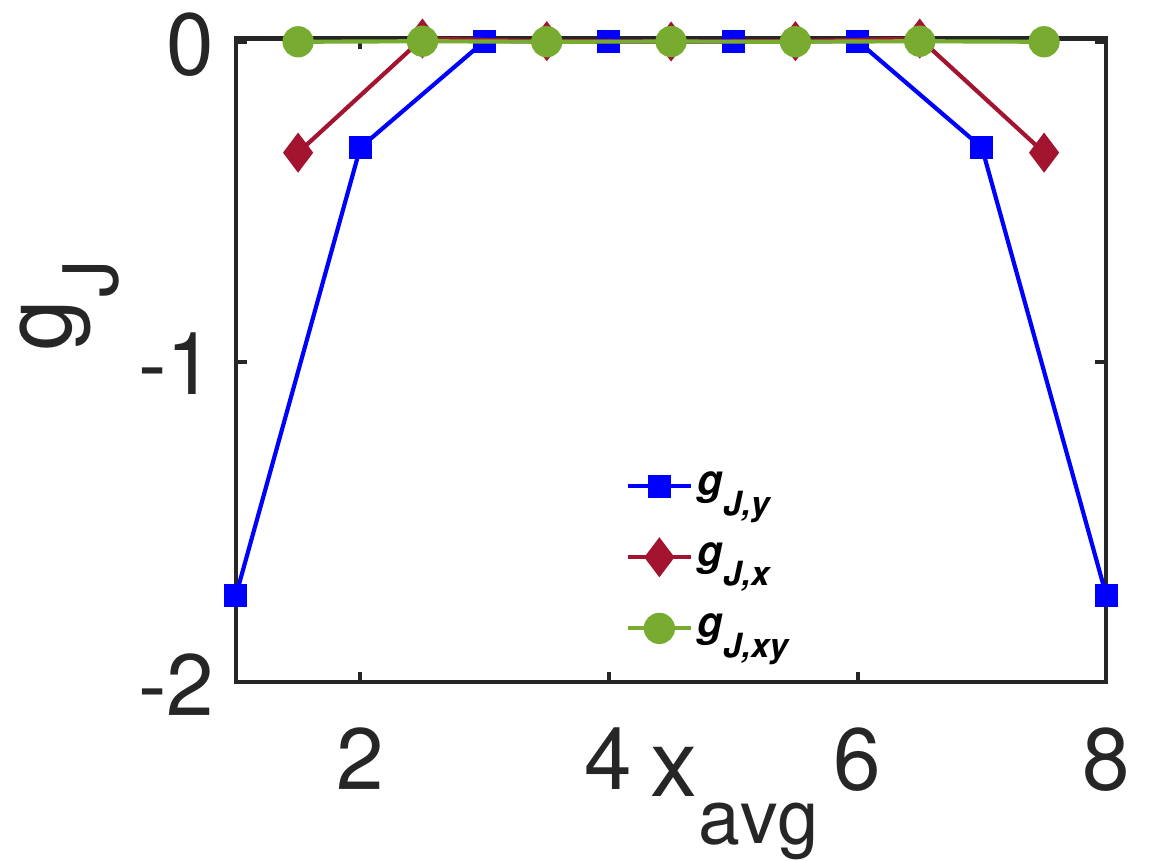}%
  \label{fig_SM_1b_rho_b}%
}\qquad
\subfloat[]{%
  \includegraphics[width=4.0cm]{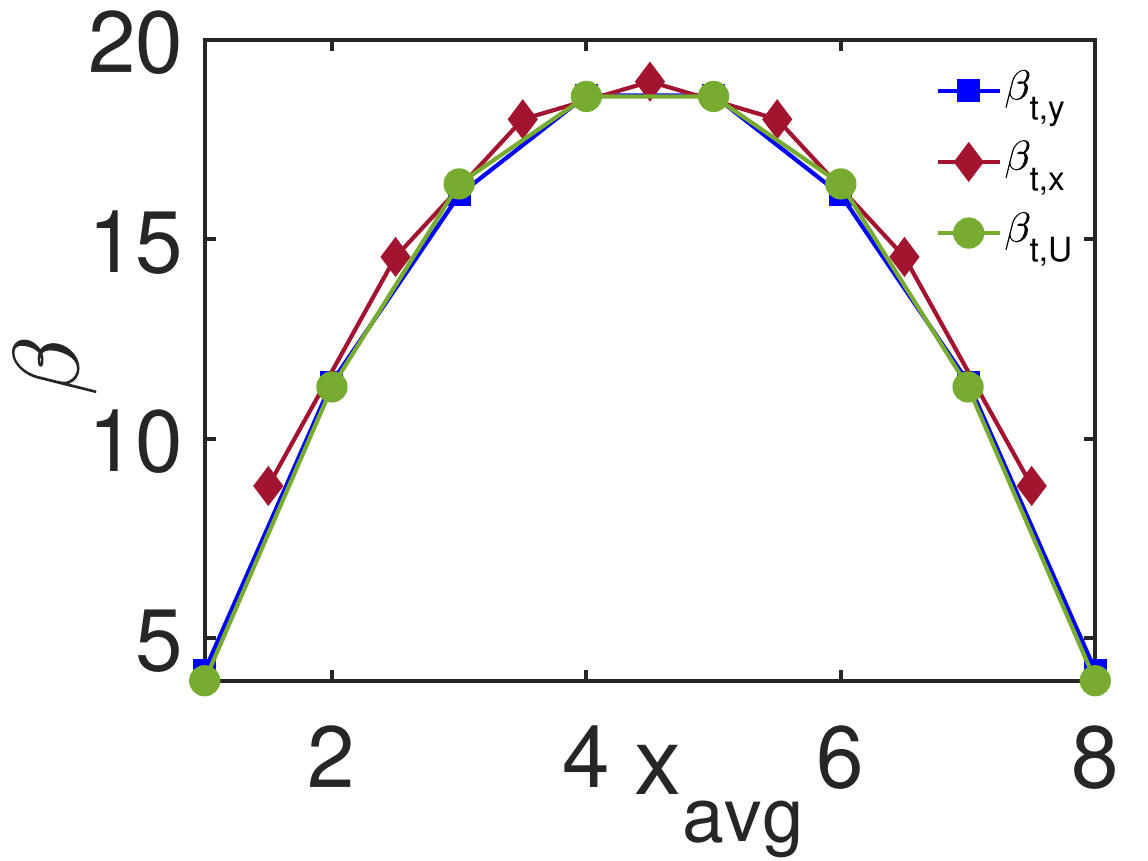}%
  \label{fig_SM_1b_RE_c}%
}
\subfloat[]{%
  \includegraphics[width=4.0cm]{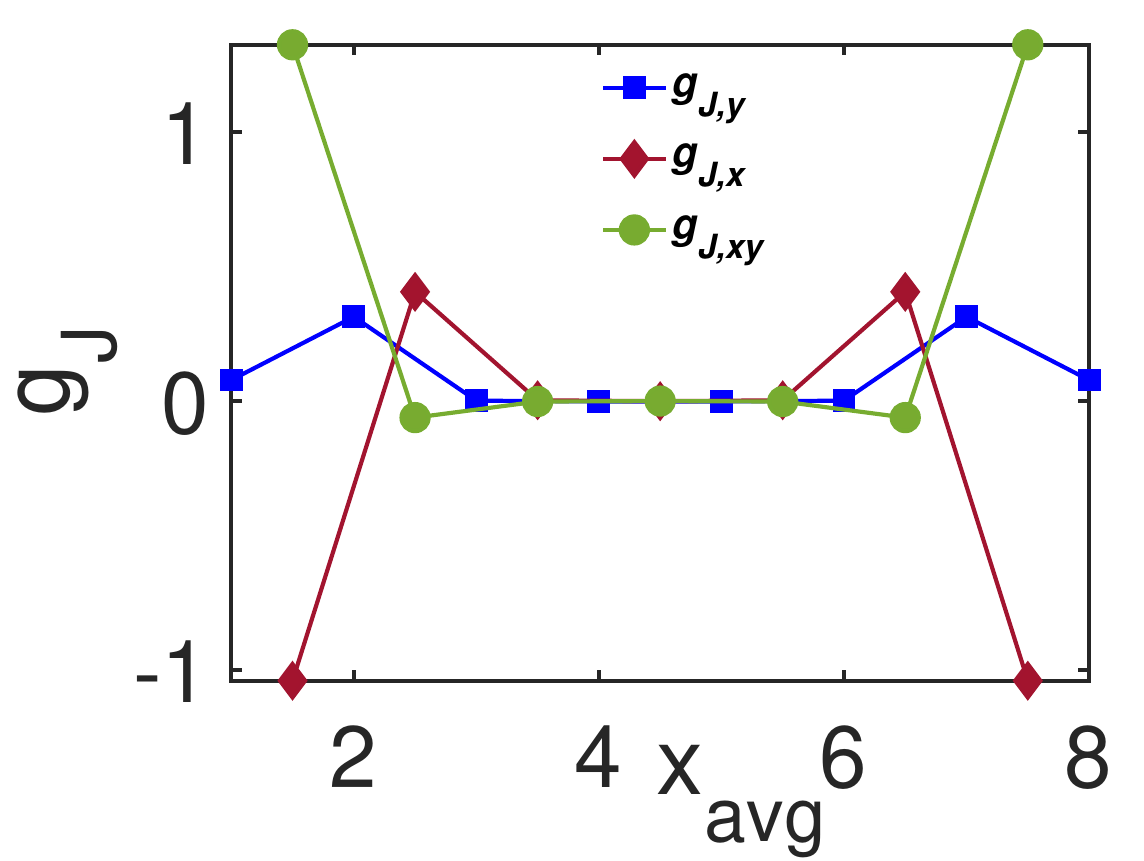}%
  \label{fig_SM_1b_rho_d}%
}
\caption{\raggedright Same as Fig.~\ref{fig_SM_1b_GF} but for $\Delta_3$ cost function (RDM distance).  Likewise, (a) and (b) are achieved by considering $\chi=2^{11}$, while (c) and (d) by $\chi=2^9$.} 
\label{fig_SM_1b_rho}
\end{figure}

\subsection{C. Local temperature ansatz and the initial guess for couplings}
\label{SM_C}

Now, let us assume we study the following Hamiltonian:

\bea
H = \sum_{\alpha} J_{\alpha} \hat{O}_{\alpha},
\eea
where, due to the locality of the Hamiltonian, only certain $J_{\alpha}$'s are nonzero. We are interested in finding the second quantization form of the EH expanded as follows:
\bea
K_A = \sum_{\alpha \in A} g_{\alpha} \hat{O}_{\alpha}.
\eea
Here, due to the renormalization procedure involved in tracing the degrees of freedom outside $A$, $g_{\alpha}$'s can be viewed as our running coupling constants which $J_{\alpha}$ has flown to. Thus, in principle, any $g_{\alpha}$ consistent with symmetry considerations emerge. In practice, only a small set of them will be relevant and non-negligible.

In our algorithm, we are trying to find $g_{\alpha}$ numerically, assuming (a subset of relevant) correlation functions are known. In our optimization algorithm, we initialized the coupling constants of the EH, $g_{\alpha}$, using LTA's ideal form. In LTA, the EH is local and its coupling constants, $g_{\alpha}$'s, are nonzero only when the corresponding couplings of the Hamiltonian (UV theory), $J_{\alpha}$'s, are nonzero. Another task in LTA is to assign a position to each operator. For simple two-point operators (such as $\bf S_i.S_j$ in the Heisenberg model, or $c_{\bf i,\sigma}^\dag c_{\bf j,\sigma}$ in the Hubbard model), the position is defined as the average position of its components, namely $\overline{\bf ij} = \frac{\bf i+j}{2}$. Next, we must compute the minimum distance (geodesics) between $\overline{\bf ij}$ and the boundary separating $A$ and its environment, $B$. Let us call this minimum distance, $x_{\bf ij}$. Finally, at zero temperature (for ground-states) and for the open boundary condition (OBC), LTA attributes the following form to $g_{\alpha}\para{x_{\bf ij}}$~\cite{cardy2016entanglement}:

\begin{figure}[t]
\centering
\includegraphics[width=8.0cm]{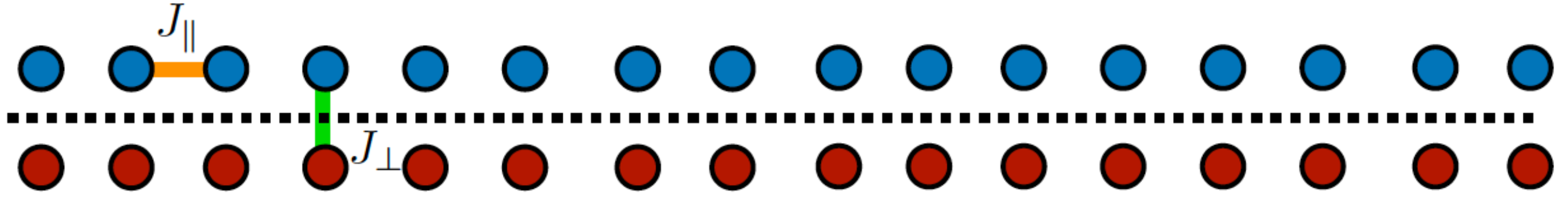}%
\caption{\raggedright We study the Heisenberg model on this ladder for $J_{\perp} = 0.5$, and $J_{\parallel} = 1$. Subsystem $A$, whose EH is desired, is denoted by blue sites.} 
\label{fig_SM_0c}
\end{figure}

\begin{figure}[t]
\centering
\subfloat[]{%
  \includegraphics[width=4.0cm]{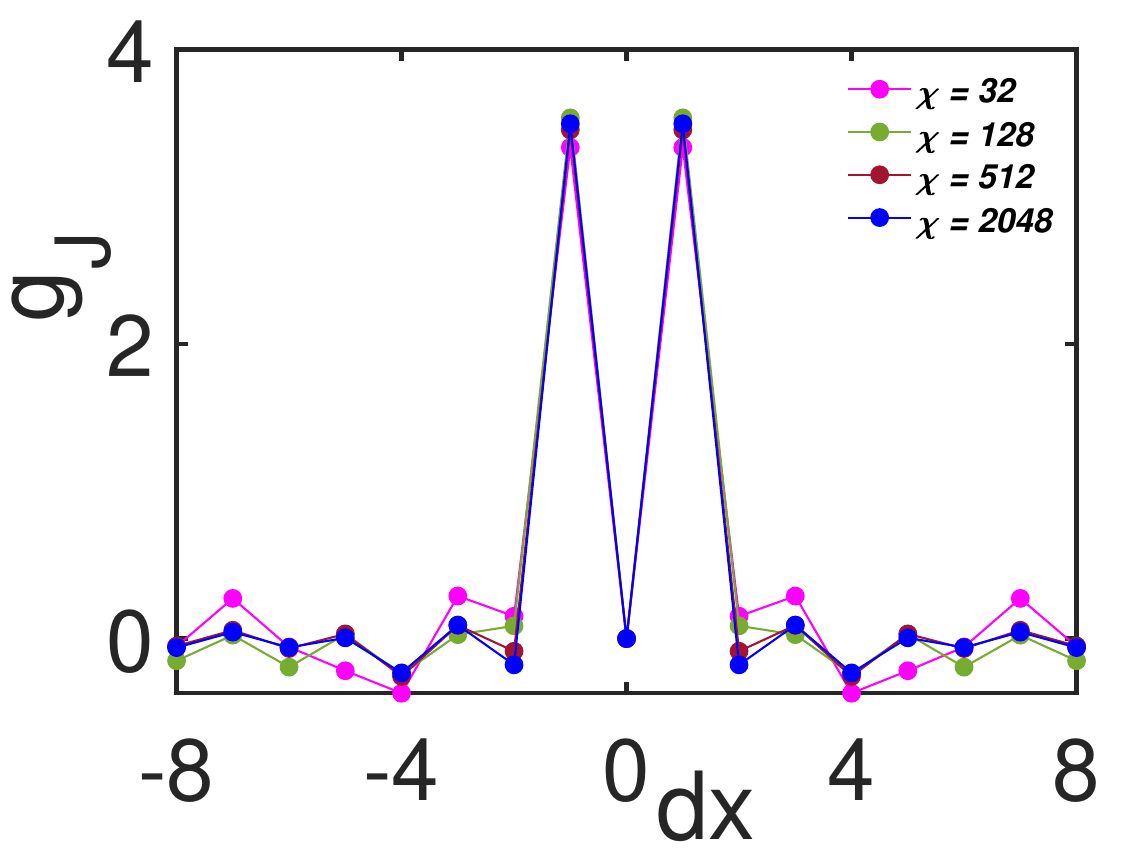}%
  \label{fig_SM_2a_a}%
}
\subfloat[]{%
  \includegraphics[width=4.0cm]{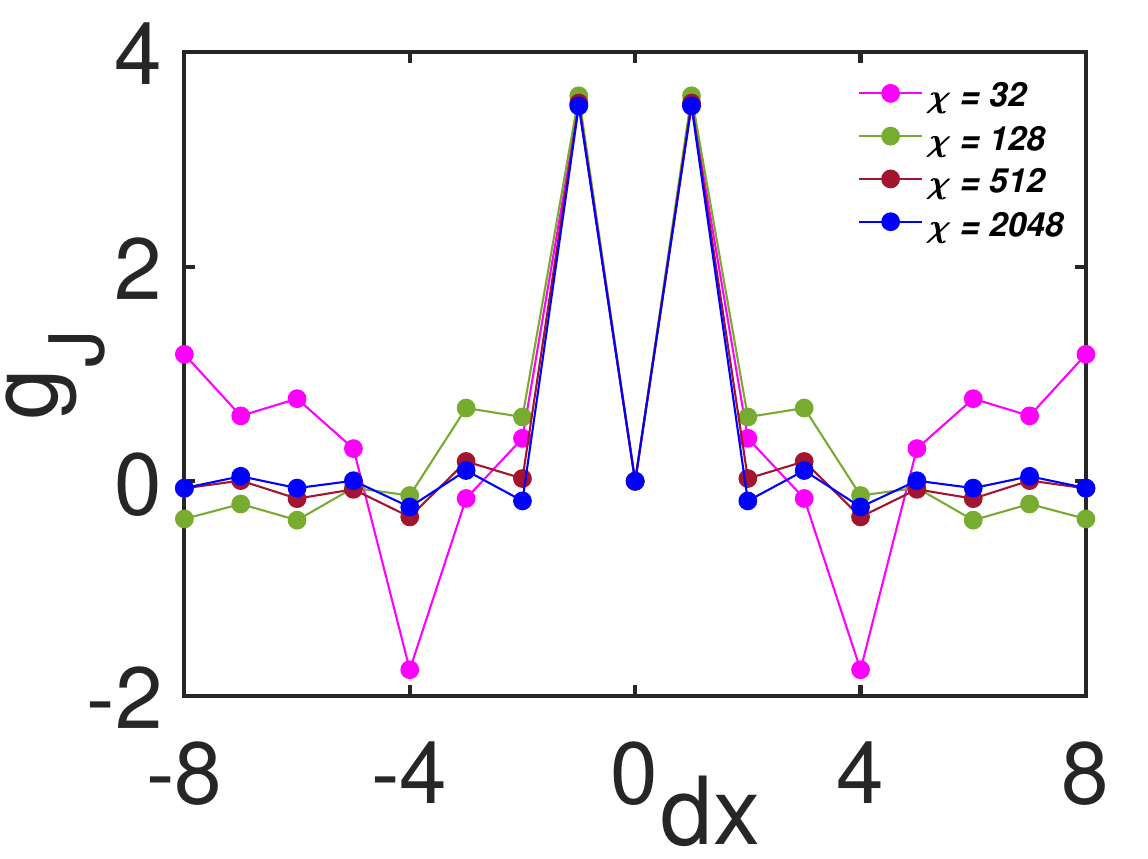}%
  \label{fig_SM_2a_b}%
}
\caption{\raggedright $K_A$'s couplings, for the geometry illustrated in Fig.~\ref{fig_SM_0c}, obtained via applying $\Delta_1$ and $\Delta_2$ cost functions and for $\chi = 2^{5}, 2^{7}, 2^{9}, 2^{11}$. The couplings are translationally invariant due to the geometry of $A$. (a) $g_{J,dx}$ obtained by minimizing $\Delta_1$ cost function. (b) $g_{J,dx}$ obtained by minimizing $\Delta_2$ cost function. As we see, both cost functions result in the same couplings for $\chi = 2^{11}$. Also, $\Delta_1$ results exhibits much less fluctuations than $\Delta_2$'s and thus are more reliable for smaller bond dimensions ($\chi$'s).} 
\label{fig_SM_2a}
\end{figure}

\bea
g_{\alpha}\para{x_{\bf ij}} = J_{\alpha} \frac{4\ell}{v} \sin \para{\frac{\pi}{2\ell} x_{\bf ij}}.
\eea
where $\ell$ is the maximum value of $x_{\bf ij}$ (i.e., the linear dimension of $A$ normal to $\partial A$), and $v$ is the group velocity of low energy excitations (quasi-particles) and is model-dependent. In our algorithm, besides $v$, we also treated $\ell$ as a variational parameter. 
We first optimized and tuned $v$, and $\ell$. Then, we took the optimized form of local $g_{\alpha}$ (associated with $v^*$, and $\ell^*$), and using the gradient descent algorithm we optimized our cost function. We allowed all relevant couplings, including distant neighbors and non-local terms (which were absent in the system's Hamiltonian) as well as the initialized local terms to vary and deviate from their initial point. Therefore, we have not imposed locality in our procedure, although it finally emerged naturally as the optimum solution (except at the boundary of $A$ with $B$, where farther neighbors became more pronounced).

Similarly, for the periodic boundary condition (PBC) at $T=0$, LTA assigns the following value to $g_{\alpha}\para{x_{\bf ij}}$~\cite{cardy2016entanglement}:
\bea
g_{\alpha}\para{x_{\bf ij}} = J_{\alpha} \frac{2 L}{v} \frac{ \sin \para{\frac{\pi}{L} x_{\bf ij}} \sin \para{\frac{\pi}{L} \para{\ell -x_{\bf ij}}}}{\sin \para{\frac{\pi}{L} \ell}},
\eea 
where $L$ is size of the entire system ($M$) in the direction normal to $\partial A$.

\subsection{D. A detailed comparison between the performance of $\Delta_1$, $\Delta_2$ and $\Delta_3$ cost functions}
\label{SM_D}

Here, we compare the EH's coefficients obtained by utilizing all three cost functions for the ladder geometry and for the Hubbard and Heisenberg models. 

\begin{figure}[t]
\centering
\includegraphics[width=8.0cm]{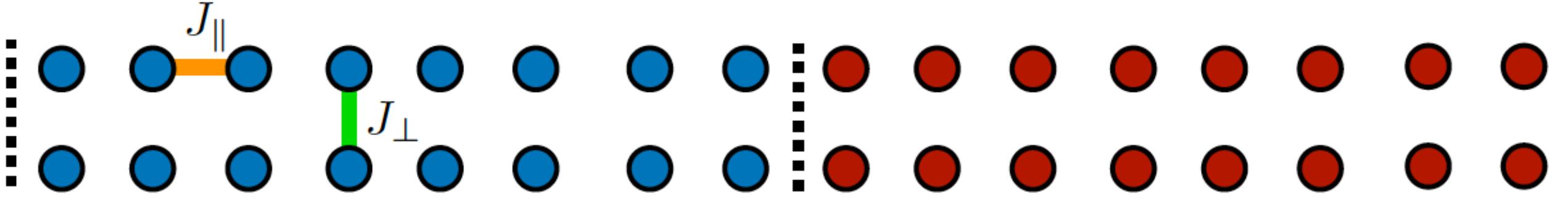}%
\caption{\raggedright We study the Heisenberg model on this ladder for $J_{\perp} = 2$, and $J_{\parallel} = 1$. Subsystem $A$, whose EH is desired, is denoted by blue sites.} 
\label{fig_SM_0d}
\end{figure}

\begin{figure}[t]
\centering
\subfloat[]{%
  \includegraphics[width=4.0cm]{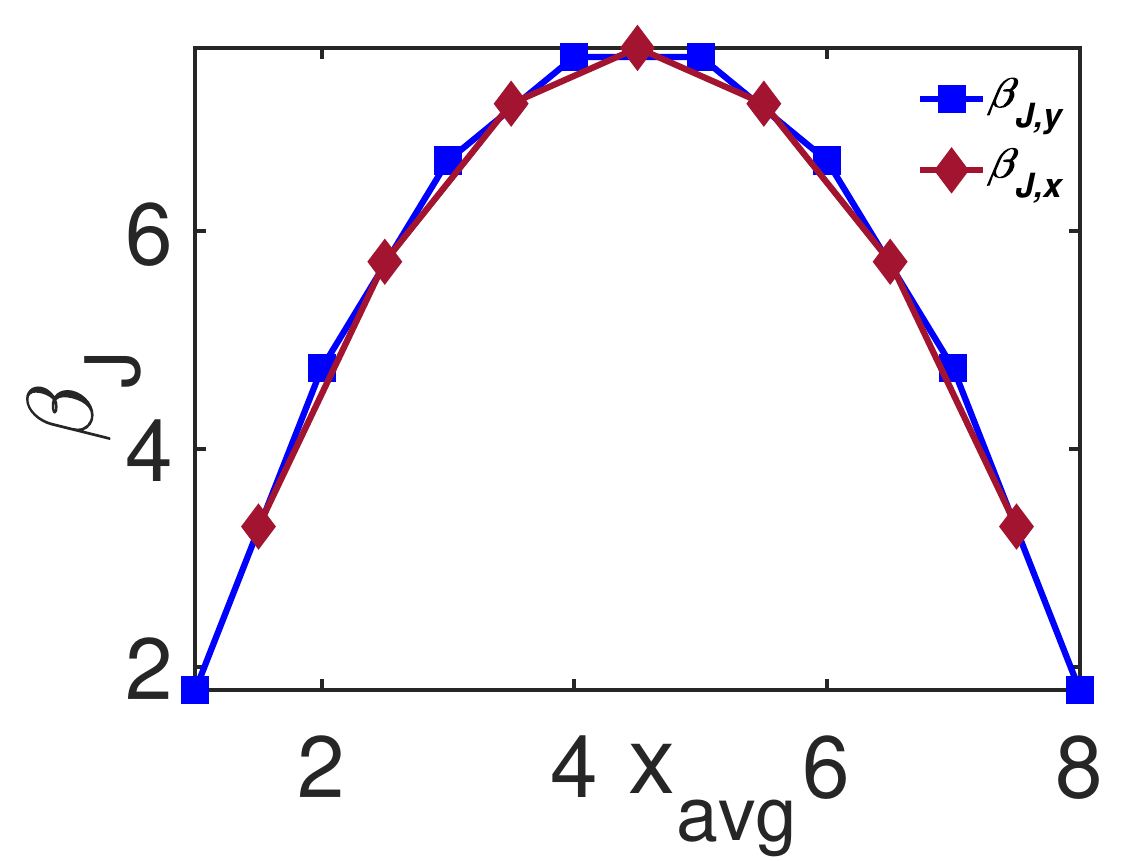}%
  \label{fig_SM_2b_GF_a}%
}
\subfloat[]{%
  \includegraphics[width=4.0cm]{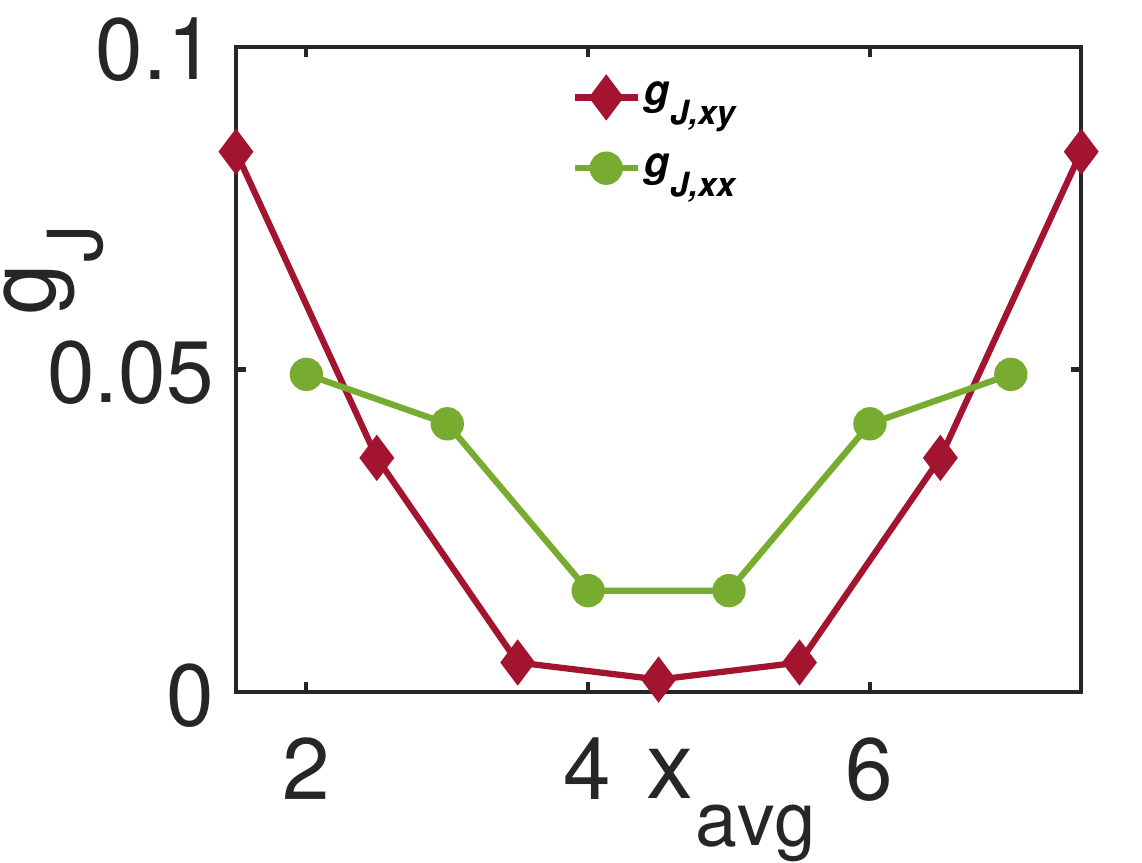}%
  \label{fig_SM_2b_GF_b}%
}\qquad
\subfloat[]{%
  \includegraphics[width=4.0cm]{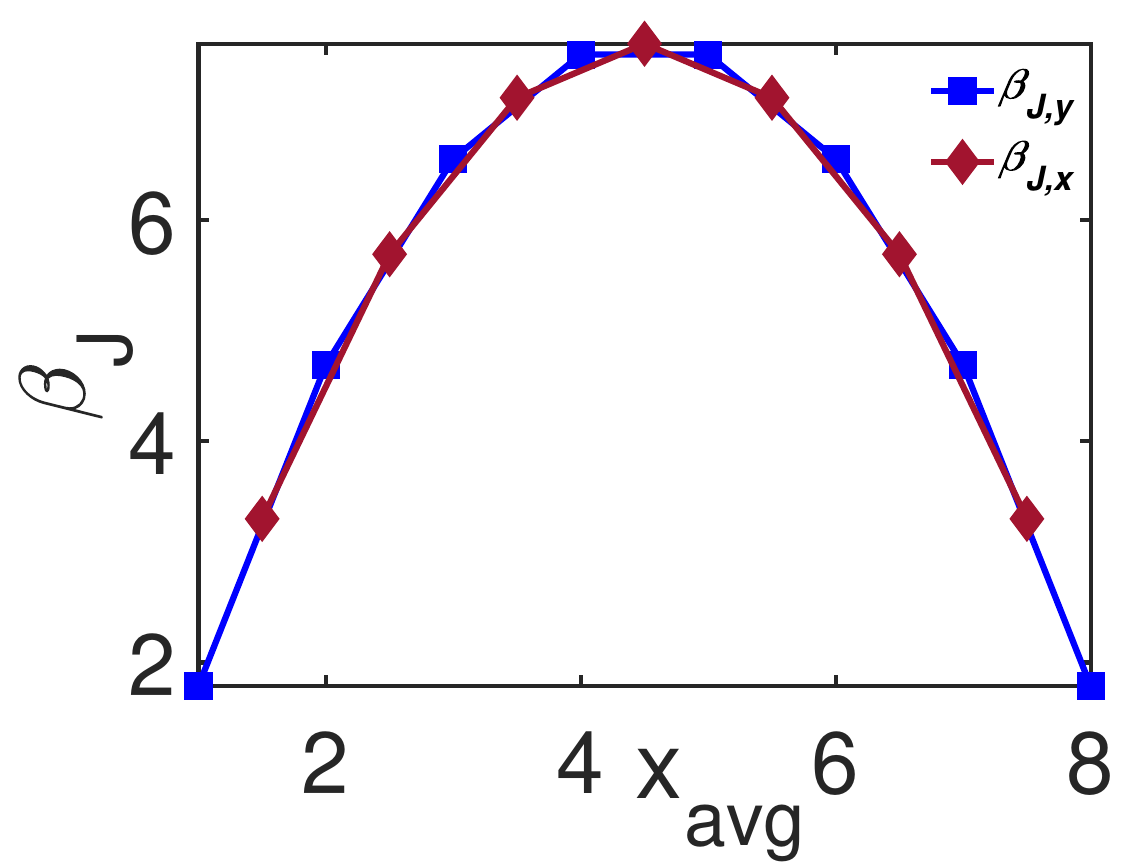}%
  \label{fig_SM_2b_GF_e}%
}
\subfloat[]{%
  \includegraphics[width=4.0cm]{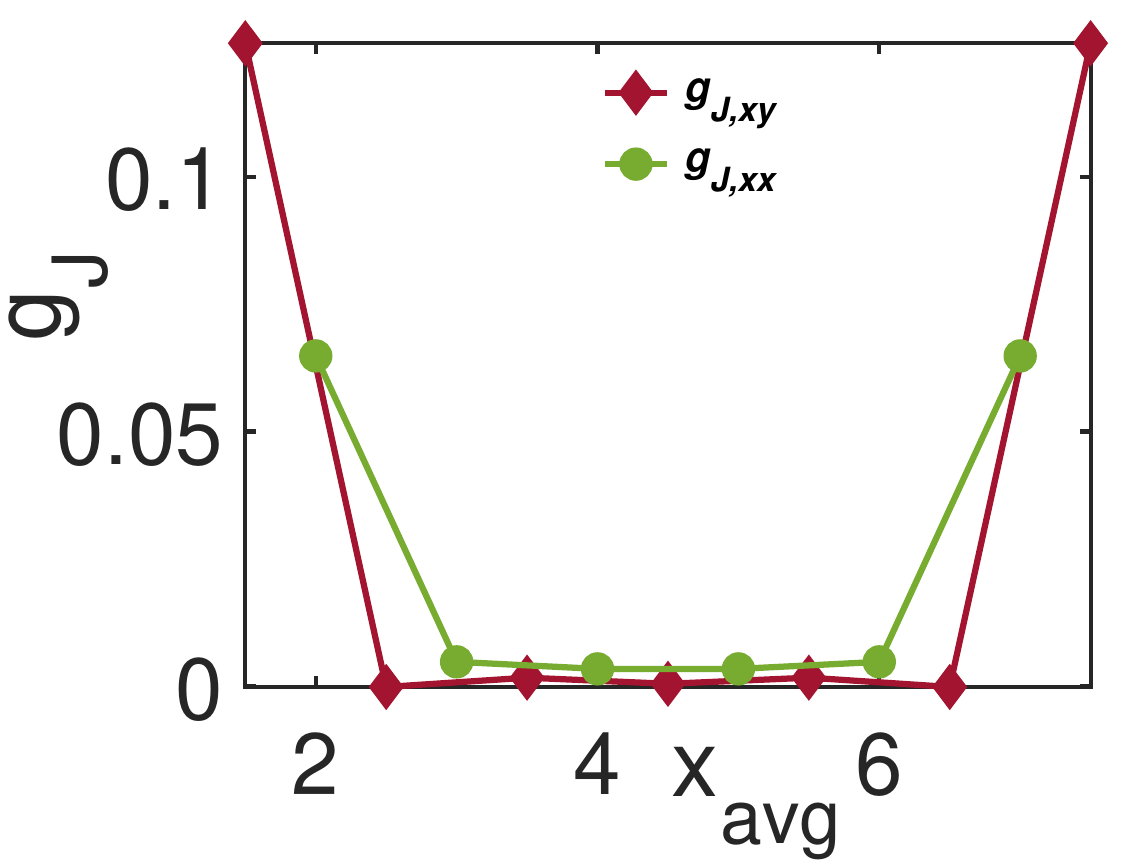}%
  \label{fig_SM_2b_GF_f}%
}
\caption{\raggedright $K_A$'s couplings, for the geometry illustrated in Fig.~\ref{fig_SM_0c}, obtained via applying $\Delta_1$ cost functions and for $\chi = 2^{7}, 2^{11}$.(a) and (c) show the inverse local temperatures ($\beta_y\para{i_x}:= \frac{1}{J_{\perp}}g_{J,i_x,i_x}$, $\beta_x\para{i_x+1/2}:= \frac{1}{J_{\parallel}}g_{J,i_x,i_x+1}$) for $\chi = 2^{11}$, and $2^7$, respectively. (b) and (d) present the second and third neighbor couplings of the EH.} 
\label{fig_SM_2b_GF}
\end{figure}

\begin{figure}[t]
\centering
\subfloat[]{%
  \includegraphics[width=4.0cm]{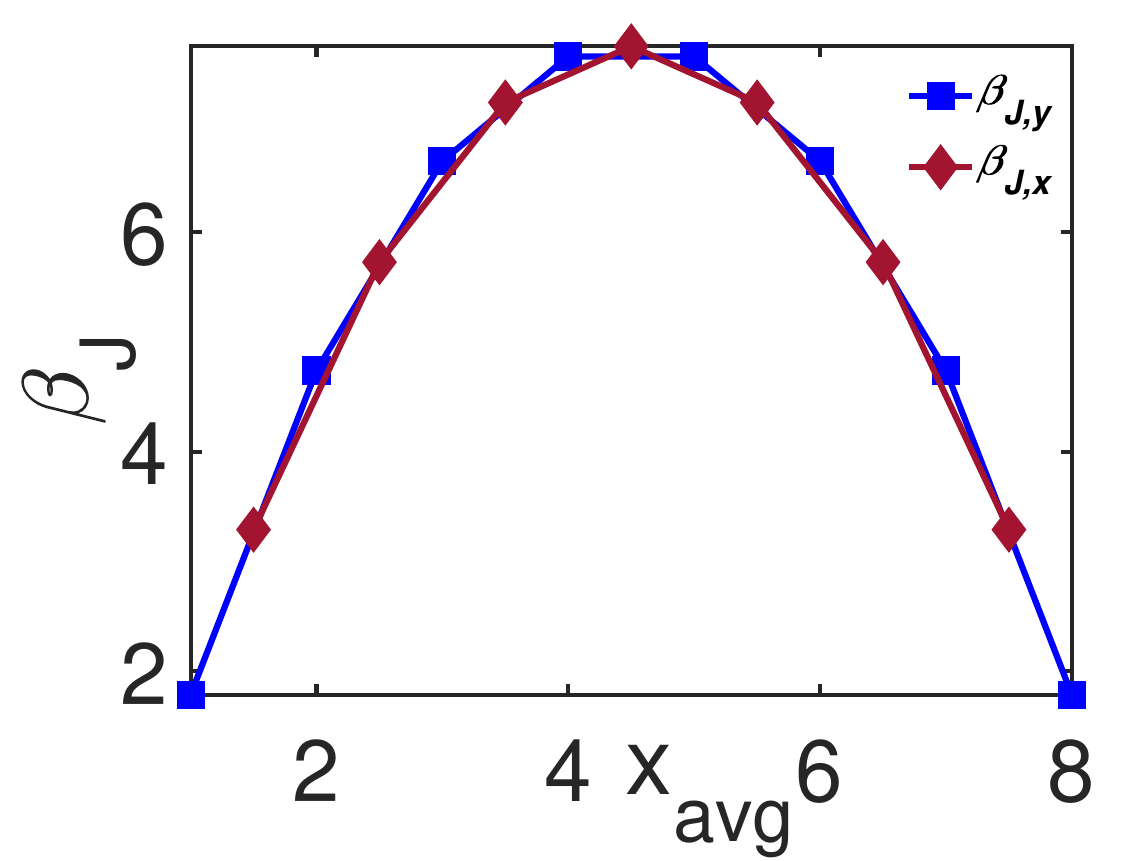}%
  \label{fig_SM_2b_RE_a}%
}
\subfloat[]{%
  \includegraphics[width=4.0cm]{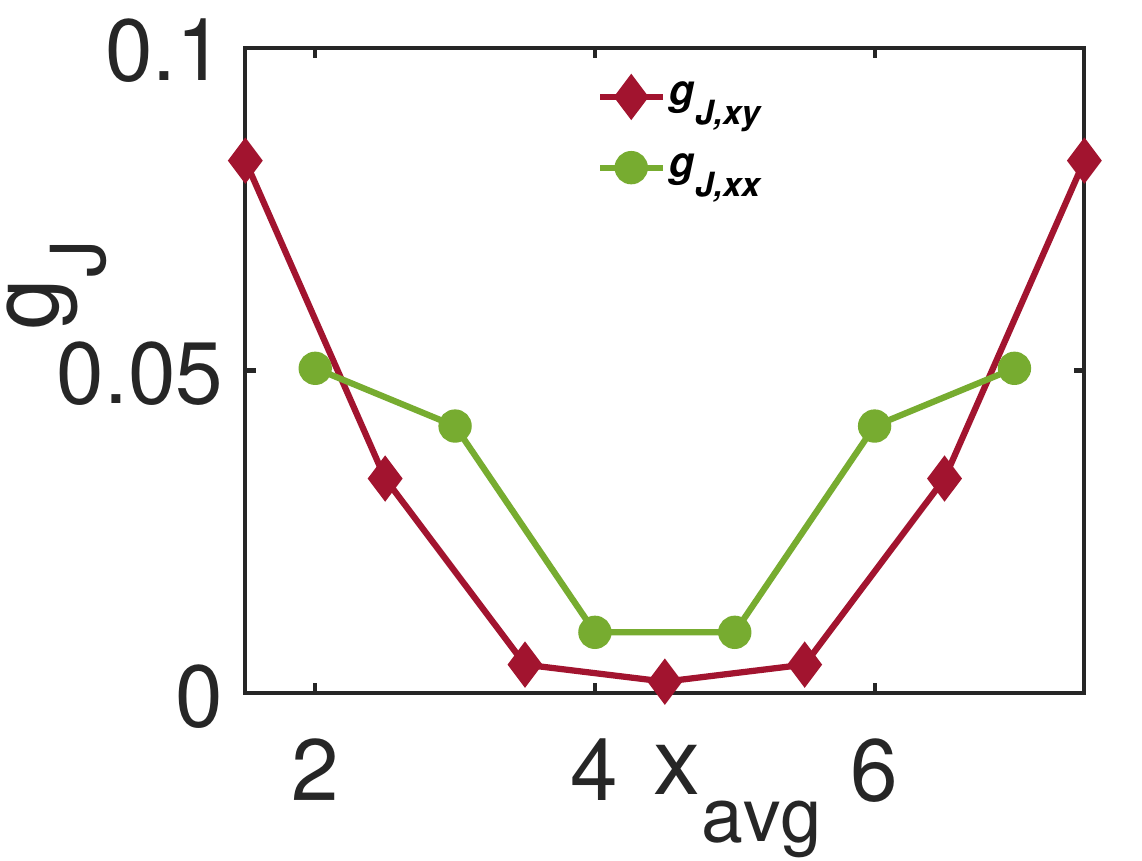}%
  \label{fig_SM_2b_RE_b}%
}\qquad
\subfloat[]{%
  \includegraphics[width=4.0cm]{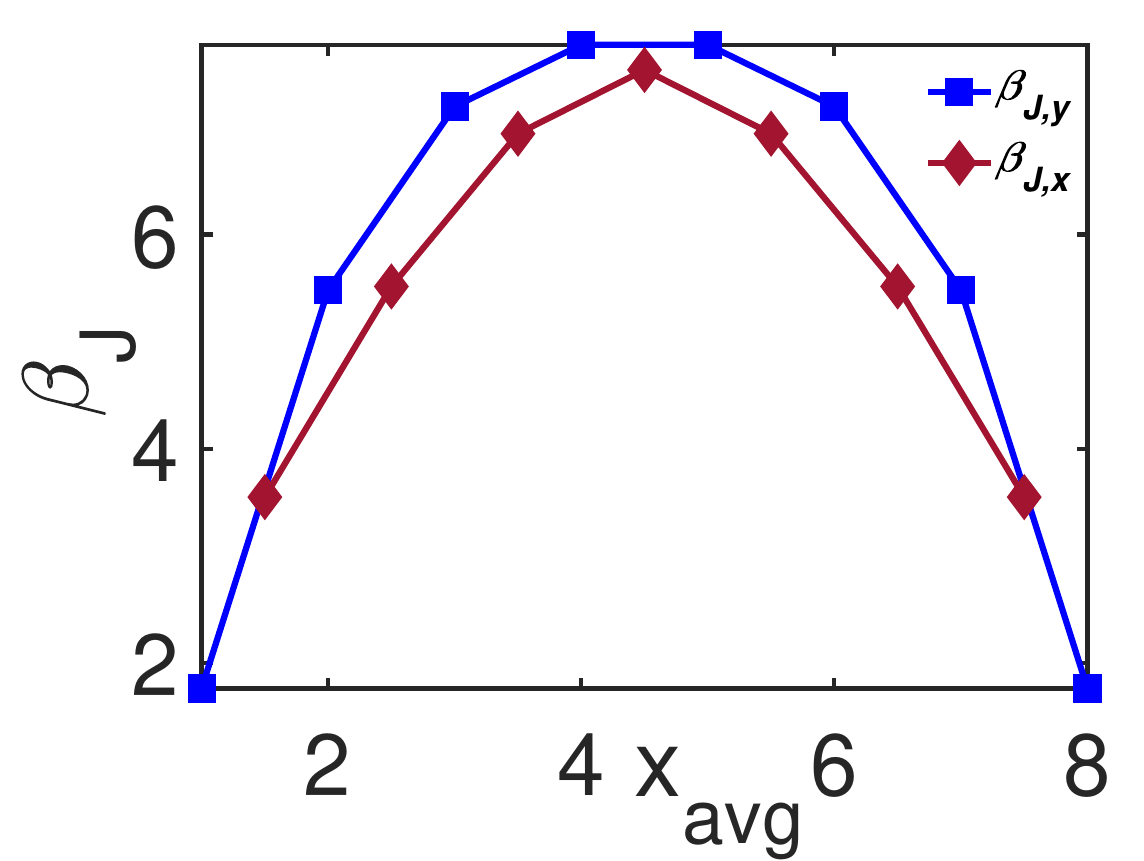}%
  \label{fig_SM_2b_RE_e}%
}
\subfloat[]{%
  \includegraphics[width=4.0cm]{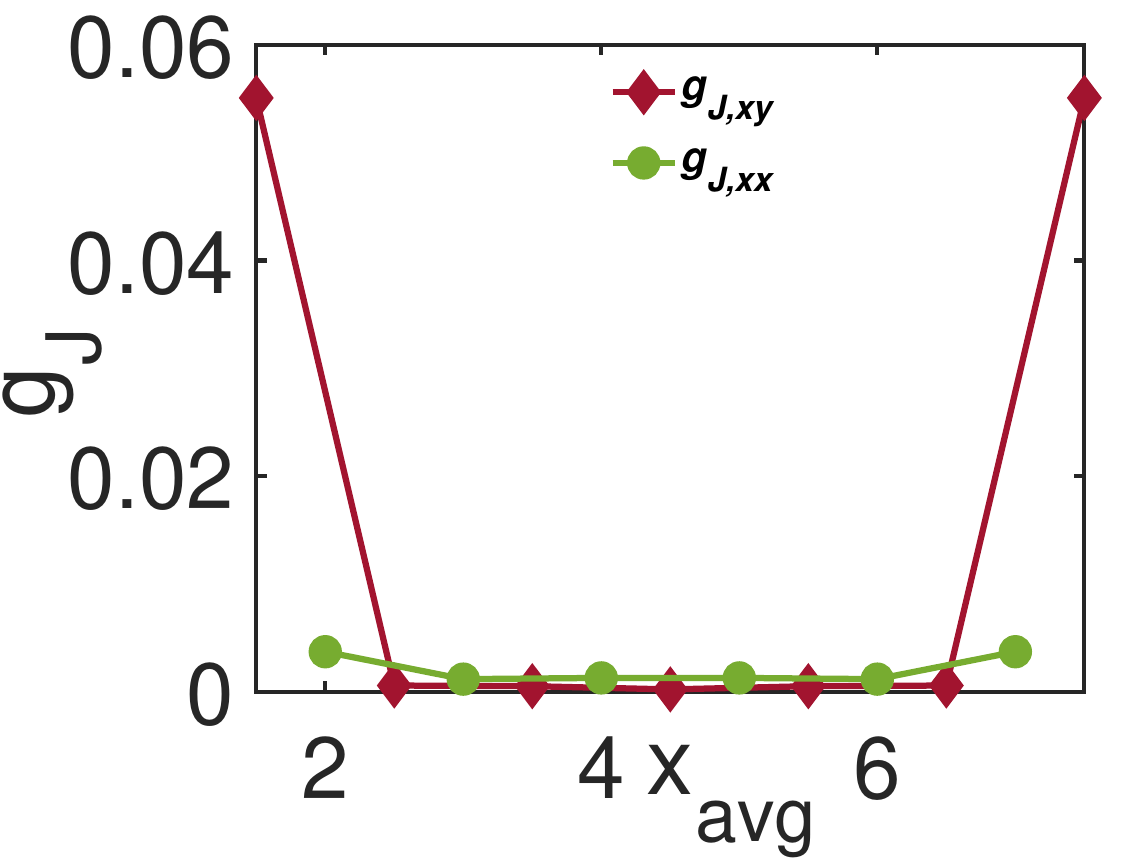}%
  \label{fig_SM_2b_RE_f}%
}
\caption{\raggedright Same as Fig.~\ref{fig_SM_2b_GF} but for $\Delta_2$ cost function. (a) and (b) are obtained by keeping $\chi=2^{11}$ basis states of the Hilbert space in DMRG, while (c) and (d) by $\chi=2^7$.} 
\label{fig_SM_2b_RE}
\end{figure}

We first consider the undoped Hubbard model on a ladder geometry (in which $U=4$, $t_{\perp} = 0.5$, $t_{\parallel}=1$) depicted in Fig.~\ref{fig_SM_0a}. This geometry results in a highly entangled subsystem, indeed a volume law entanglement entropy. We apply all three cost functions to this system for the following five different bond dimensions: $\chi = 2^5, 2^7, 2^9, 2^{11}$. The EH for this case is translationally invariant, namely $g_{\tau,i_x,j_x} = g_{\tau,dx_{ij}}$ ($\tau=t,U,J,V$), where $dx_{ij}:=j_x-i_x$. In this section, the translational symmetry is imposed on the couplings explicitly. We first compare the EH's couplings achieved via applying the GF distance ($\Delta_1$), QRE ($\Delta_2$), and the RDM distance  ($\Delta_3$) for $\chi= 2^{11}$ (see Fig.~\ref{fig_SM_1a}). With this bond dimension, we can nearly probe the ground-state properties. The coefficients of local terms in the EH are almost consistent in these three methods. On the other hand, we know that ideally the ground-state must exhibit particle-hole symmetry. Although $\chi = 2^{11}$ is still insufficient for true convergence in DMRG for such a highly entangled state ($\chi = 2^{12}$ seems to be enough), $\Delta_1$'s results reflect the particle-hole symmetry (e.g., the (renormalized) second neighbor hoppings are infinitesimal), while for those of $\Delta_2$ and $\Delta_3$ the particle-hole symmetry is visibly violated due to overfitting issues mentioned previously. Furthermore, a previous quantum Monte Carlo based study of a similar situation~\cite{toldin2018entanglement} indicated the irrelevance of $g_V$ couplings which is consistent with $\Delta_1$'s estimations. 
Additionally, perturbative studies of the EH indicate an oscillating spin-spin couplings~\cite{furukawa2011entanglement,lauchli2012entanglement,chen2013quantum} (though subdominant to the $\it renormalized$ onsite interaction) which agrees well with our results via minimizing GF distance ($\Delta_1$), while those of $\Delta_2$ and $\Delta_3$ exhibit deviations in addition to their overestimation for the spin-spin couplings. In Fig.~\ref{fig_SM_1a_d}, we plot the normalized onsite interaction strength $U_{\rm eff}^{(\chi)} := \frac{g^{(\chi)}_U}{g^{(\chi)}_t\para{1}}$ for all four bond dimensions considered in our investigations. Again, as we see in Fig.~\ref{fig_SM_1a_d}, the results of the GF distance ($\Delta_1$) are more robust and less sensitive to $\chi$, despite several orders of magnitude change in $\chi$, while those of the QRE and RDM distance display stronger fluctuations.

Now, we turn to the geometry shown in Fig.~\ref{fig_SM_0b} (where $U=4$, $t_{\perp} = 2$, $t_{\parallel}=1$) and present our results for all three cost functions in Figs.~\ref{fig_SM_1b_GF},~\ref{fig_SM_1b_RE}, and \ref{fig_SM_1b_rho}. Here, we have defined the following inverse local temperatures: $\beta_{t,x}\para{i_x+1/2} := g_{t,\bf i,i + \hat{x}}$, $\beta_{t,y}\para{i_x} := g_{t,\bf i,i + \hat{y}}$, and $\beta_{U}\para{i_x} := \frac{1}{U}g_{U,\bf i}$. For this problem, due to the PBC imposed along $x$ direction, we found out that even with $\chi = 2^{11}$, there is still some room for DMRG to converge to the true ground-state. As a result, we still see some minor discrepancy among the results of the three methods for $\chi = 2^{11}$ for subdominant and correction terms beyond LTA (though they yield highly correlated results). Nonetheless, the local terms (i.e., dominant couplings) are reasonably consistent. We have also plotted the results of $\chi = 2^9$ for all three methods and again, $\Delta_1$'s results proved to be more robust and $\Delta_2$ and $\Delta_3$'s less stable. Thus, in the presence of truncation errors, we can trust the results of the GF distance more than those of the other two candidates for the cost function. 

For the sake of completeness, we have also explored the robustness and the accuracy of the above three cost function candidates for the Heisenberg model, again on a ladder geometry. To this end, we first studied the geometry shown in Fig.~\ref{fig_SM_0c} (where $J_{\perp} = 0.5$, $J_{\parallel}=1$), and presented its results in Fig.~\ref{fig_SM_2a}. Similar to the Hubbard model case, leads to a highly entangled ground-state. Likewise, we expect translationally invariant couplings, namely $g_{J,i_x,j_x} = g_{J,dx_{ij}}$. We have presented $g_{J,dx}$ for $\chi = 2^5, 2^7, 2^9, 2^{11}$ for the GF and QRE cost functions (the RDM cost function yields results similar to that of the QRE). In this case as well, the GF distance turns out to be the most stable one.

Finally, we studied the Heisenberg model on the geometry illustrated in Fig.~\ref{fig_SM_0d} (in which $J_{\perp} = 0.5$, $J_{\parallel}=1$). Their results are presented in Figs.~\ref{fig_SM_2b_GF} and \ref{fig_SM_2b_RE} for the GF distance and QRE, respectively. For this problem, we indeed achieved the true ground-state using $\chi=2^{11}$. Therefore, all cost functions must achieve the same couplings. On the other hand, for smaller bond dimensions, e.g., $\chi = 2^7$, $\Delta_1$ achieves more accurate results (relative to $\chi = 2^{11}$) than the remaining cost functions.
\bibliographystyle{unsrtnat}
\putbib
\end{bibunit}
\end{document}